\documentclass[draft,tightenlines,nofootinbib,preprint,aps,eqsecnum,amsmath,amssymb]{revtex4}

\newcommand{\beq}{\begin{equation}}
\newcommand{\eeq}{\end{equation}}
\newcommand{\bea}{\begin{eqnarray}}
\newcommand{\eea}{\end{eqnarray}}
\newcommand{\cir}{{\buildrel \circ \over =}}
\newcommand{\sgn}{\epsilon}

\newcommand{\A}{{\dot x}^2}
\newcommand{\B}{{x'}^2}
\newcommand{\C}{(\dot x\cdot x')}
\newcommand{\varb}{(\tau ,\lambda)}

\newcommand{\pri}{'}
\newcommand{\xpu}{\dot x}
\newcommand{\emu}{^{\mu}}
\newcommand{\enu}{^{\nu}}

\begin{document}

\title{The Rest-Frame Instant Form and Dirac Observables for the Open Nambu String}

\medskip

\author{David Alba}

\affiliation{Dipartimento di Fisica\\
Universita' di Firenze\\Polo Scientifico, via Sansone 1\\
 50019 Sesto Fiorentino, Italy\\
 E-mail ALBA@FI.INFN.IT}

\author{Horace W. Crater}

\affiliation{
The University of Tennessee Space Institute \\
Tullahoma, TN 37388 USA \\ E-mail: hcrater@utsi.edu}

\author{Luca Lusanna}

\affiliation{ Sezione INFN di Firenze\\ Polo Scientifico\\ Via Sansone 1\\
50019 Sesto Fiorentino (FI), Italy\\  E-mail: lusanna@fi.infn.it}

\begin{abstract}

The rest-frame instant form of the positive-energy part of the open
Nambu string is developed. The string is described as a decoupled
non-local canonical non-covariant Newton-Wigner center of mass plus
a canonical basis of Wigner-covariant relative variables living in
the Wigner 3-spaces. The center of mass carries a realization of the
Poincare' algebra depending upon the invariant mass and the
rest-spin of the string, functions of the relative variables. A
canonical basis of gauge invariant Dirac observables is built with
Frenet-Serret geometrical methods. Some comments on canonical
quantization are made.

\end{abstract}

\today

\maketitle

\vfill\eject

\section{Introduction}

The recently developed {\it parametrized Minkowski theories}
\cite{1a,2a,3a} allow one to give a unified description of isolated
relativistic systems (particles, strings, fluids, fields) admitting
a Lagrangian description both in inertial and non-inertial frames of
Minkowski space-time. A basic tool is the family of admissible 3+1
splittings of Minkowski space-time \cite{3a} (i.e. a convention for
clock synchronization identifying the instantaneous 3-spaces
\cite{2a}) and the use of radar 4-coordinates adapted to a time-like
observer. In this way it is possible to show that the transition
from a non-inertial frame to every other frame is a {\it gauge
transformation}: as a consequence special relativistic physics is
not influenced by the choice of the clock synchronization convention
and by the choice of the 3-coordinates on the instantaneous 3-space
(only the appearance of the phenomena is changed).\medskip

In particular we can define the {\it inertial rest-frame instant
form of dynamics} of every isolated system by considering a 3+1
splitting whose instantaneous 3-spaces are  orthogonal to the
conserved 4-momentum of the isolated system: it corresponds to the
intrinsic inertial rest frame centered on an inertial observer
\footnote{See Ref.\cite{3a} for the non-inertial rest frames and for
the associated non-inertial rest-frame instant form of dynamics.}.
In this formulation there is a complete understanding and
classification \cite{4a,5a} of the collective relativistic
4-variables (canonical non-covariant Newton-Wigner center of mass,
non-canonical covariant Fokker-Price center of inertia,
non-canonical non-covariant M$\o$ller center of energy) for any
isolated system. They can be built by using {\it only} the ten
Poincare' generators of the isolated system (since the generators
are {\it non-local} quantities knowing the whole instantaneous
3-space, also the collective variables are non-local quantities). It
turns out that every isolated system can be described as a decoupled
non-covariant point particle (corresponding to the Newton-Wigner
center of mass) carrying a pole-dipole structure, i.e. the invariant
mass $M$ and the rest spin ${\vec {\bar S}}$ of the system.
Associated with this non-covariant non-local \footnote{No one can
observe it, so that its non-covariance is irrelevant. It is a notion
like the wave function of the universe.} particle there is an {\it
external} realization of the Poincare' algebra (its Lorentz boosts
induce Wigner rotations on Wigner spin-1 3-vectors inside the
instantaneous Euclidean 3-spaces, named Wigner 3-spaces). Inside
these Wigner 3-spaces the dynamics of the isolated system is
described by Wigner-covariant relative variables. The quantities $M$
and ${\vec {\bar S}}$ are functions only of these relative variables
and moreover they are the energy and the angular momentum of a {\it
unfaithful inner} realization of the Poincare' algebra, whose
generators are determined by the energy-momentum tensor of the
system. It is unfaithful because the inner 3-momentum vanishes: this
is the rest-frame condition identifying the rest frame. The
(interaction-dependent) inner Lorentz boosts depend on the inner
3-center of mass, canonically conjugate to the vanishing inner
3-momentum, inside the 3-space. In order not to have a double
counting of the center of mass (external, inner) we must eliminate
this variable by means of a statement on the inner Lorentz boosts.
If we ask for the vanishing of the inner Lorentz boost, it turns out
that the inner center of mass is at the origin of each instantaneous
3-space and that the world-line of the inertial observer on which
the rest frame is centered is the Fokker-Price covariant center of
inertia.
\medskip

Previously we have done a detailed study of the rest-frame instant
form of the isolated system composed of N positive-energy charged
scalar particles plus the electro-magnetic field \cite{4a,5a,6a}. In
this paper we want to extend this study to the open Nambu string
\cite{7a} (the closed one could also be studied with these methods).
As a byproduct of this description we will be able to find a
complete canonical set of classical Dirac observables for the Nambu
string, a result which was till now missing (see Ref.\cite{8a} for
previous attempts).
\medskip

As a consequence, the final description of the open Nambu string
consists of a decoupled non-covariant canonical center of mass
carrying an invariant mass $M$ and a rest-frame spin ${\vec {\bar
S}}$ functions only of the Dirac observables. The final Poincare'
algebra is the external one of a point-like particle: it depends
upon $M$ and ${\vec {\bar S}}$, whose Poisson brackets are $\{ M,
{\bar S}^i \} = 0$ and $\{ {\bar S}^i, {\bar S}^j \} =
\epsilon^{ijk}\, {\bar S}^k$. In terms of the canonical basis of
Dirac observables we have a non-linear realization of an
O(3)-algebra with generators ${\bar S}^i$, like in a 3-dimensional
$\sigma$-model, plus a O(3)-scalar invariant mass $M$ (a non-linear
functional of the Dirac observables). Therefore we have an
implementation of the old approach of Ref.\cite{9a} in terms of the
rest-frame instant form of dynamics, in which the world-sheet
4-coordinates are derived quantities (in general they are not
canonical quantities, but the analogue of predictive coordinates as
shown in Ref.\cite{4a}, where one can find the relevant
bibliography).

\bigskip

In Section II we review the standard action of the open Nambu string
and some technical points connected with the fact that its endpoints
move along null curves in Minkowski space-time, so that the
coordinates of the orthogonal gauge (OG) are singular there
\cite{10a}. Then we review the reformulation of the open string
given in Ref. \cite{11a} (see Ref.\cite{12a} for the closed
string)with the needed distributions, whose expression can be found
in Appendix A.\medskip

In Section III we reformulate the open string as a reparametrized
Minkowski theory, which allows to describe the part of the spectrum
with positive energy. Only the transversality first class
constraints are present in this formulation, because the mass shell
constraints are absent due to the different canonical variables,
parametrizing the string, introduced after an admissible 3+1
splitting of Minkowski space-time.\medskip

Then in Section IV we define the rest-frame instant form of the
positive energy open string.\medskip

In Section V we identify a Frenet-Serret (FS) canonical basis
adapted to the transversality constraints by means of a
Shanmugadhasan canonical transformation. The Dirac observables turn
out to be action-angle variables. Also a set of Lorentz-scalar Dirac
observables, not depending on angles but on the FS curvature and
torsion of the open string, is identified, but the associated
canonical basis is not explicitly known.\medskip

In Section VI we show how to replace the action-angle variables with
oscillators.\medskip

In the Conclusions, after some comments on previous attempts to find
Dirac observables of the string and a consistent quantization in $D
= 4$ dimensions, we discuss which are the problems one will face in
attempting to quantize the Dirac observables of a FS canonical
basis. Even if the quantization problem is still completely open
\footnote{In general the quantization of the reduced phase space
resulting from the solution of the constraints is inequivalent from
the quantization of the original phase space and of the constraints
followed by a quantum reduction to the physical Hilbert space.}, the
classical geometrical problems are finally understood.

 \vfill\eject

\section{A Review of the Nambu Action of the Open Strings and of the Problems
Connected with the End Points}

In this Section we discuss the kinematical problems arising from the
fact that the end points of the open string move on null surfaces.
These items are usually ignored, but they are relevant for the
choice of the symplectic structure in phase space.
\bigskip

Let us consider the action for the open Nambu string \cite{7a}
($\hbar = c = 1$; the metric is $\eta_{\mu\nu} = (+---)$) in the
notation of Ref.\cite{10a} (see also Refs.\cite{8a,11a,12a})

\beq
 S = - N\, \int_{\tau_1}^{\tau_2}\, d\tau\, \int_0^{\pi}\, d\lambda\,
 \sqrt{-h(\tau ,\lambda) }, \qquad\quad L = - N\, \sqrt{-h},
 \label{2.1}
 \eeq

\noindent where $N = {1\over{2\, \pi\, \alpha'}}$, ${\dot
x}^{\mu}(\tau ,\lambda ) = \partial x^{\mu}(\tau ,\lambda )/\partial
\tau$, $x^{{'}\mu}(\tau ,\lambda ) = \partial x^{\mu}(\tau ,\lambda
)/\partial \lambda$, and

\bea
 - h &=& - \det\parallel h_{\alpha\beta}\parallel = \C2 - \A\B \geq 0,
 \nonumber \\
 &&{}\nonumber \\
 \parallel h_{\alpha\beta}\parallel &=& \parallel
 \partial_{\alpha}\, x^{\mu}\, \partial_{\beta}\, x_{\mu}\parallel =
 \left(\begin{array}{cc}\A&{\dot x}\cdot{x'}\\ {\dot x}\cdot
 {x'}&\B\end{array}\right),\quad \alpha = 0,1,\quad \partial_0 =
 \displaystyle{\partial\  \over{\partial \tau}},\quad \partial_1 =
 \displaystyle{\partial\  \over{\partial \lambda}},\nonumber \\
  \parallel h^{\alpha\beta} \parallel &=& \displaystyle{1\over
 h}\left(\begin{array}{cc}\B&- \C\\ - \C&\A\end{array}\right).
 \label{2.2}
 \eea

\noindent The condition $-h \geq 0$ means that the surface swept by
the string in the space-time is everywhere time-like or null (i.e.
it is a causal surface). The strip ${0 < \lambda < \pi}$ is mapped
in the world-sheet spanned by the string in Minkowski space, which
is described by the coordinates $x^{\mu}(\tau ,\lambda  )$ in an
arbitrary inertial frame.
\medskip

Let us define the following two quantities

\bea
 P^{\mu}(\tau ,\lambda) &=& -\displaystyle{\partial L\over{\partial
\dot x_{\mu}(\tau ,\lambda)}} = N\, \sqrt{-h}\, h^{0\alpha}\,
\partial_{\alpha}\, x^{\mu} = \displaystyle{N\over{\sqrt{-h}}}\,
\Big(\C\, {x'}^{\mu} - \B\, {\dot x}^{\mu}\Big),\nonumber \\
 \Pi^{\mu}(\tau ,\lambda) &=&
- \displaystyle{\partial L\over{\partial {x'}_{\mu}(\tau ,\lambda)}}
= N\, \sqrt{-h}\, h^{1\alpha}\, \partial_{\alpha}\, x^{\mu} =
\displaystyle{N\over{\sqrt{-h}}}\, \Big(\C\, {\dot x}^{\mu} - \A\,
{x'}^{\mu}\Big),
 \label{2.3}
 \eea

\noindent where $P^{\mu}$ is the canonical momentum, which satisfies
the identities

\bea
 &&P^2(\tau ,\lambda ) + N^2\, \B\varb = 0,\qquad P\varb \cdot x\pri\varb =
 0,\qquad \Pi^2\varb + N^2\A\varb = 0,    \nonumber \\
 &&\Pi\varb \cdot \xpu\varb = 0,\qquad \Pi\varb \cdot x\pri\varb  =
N\sqrt{-h\varb },\nonumber \\
 &&\Pi\varb \cdot P\varb  = N^2\xpu\varb \cdot {x'}\varb, \qquad P\varb \cdot
 \xpu\varb  = N\sqrt{-h\varb }.\nonumber \\
 &&{}
 \label{2.4}
 \eea

\noindent At the Hamiltonian level we get the usual first class
constraints $P^2(\tau ,\lambda ) + N^2\, \B\varb \approx 0$ and
$P\varb \cdot x\pri\varb \approx 0$.
\medskip

The Hessian matrix is \cite{10a}

\bea
 \parallel W^{\mu\nu}(\tau ,\lambda)\parallel &=&
\parallel \displaystyle{\partial^2 L\over{\partial {\dot x}_{\mu}
\, \partial {\dot x}_{\nu}}}\parallel = \nonumber \\
 &=& \parallel\displaystyle{N\, \B\over{(- h)^{3\over 2}}}\,
\Big[- h\, {\eta}^{\mu\nu} + \B\, {\dot x}^{\mu}\, {\dot x}^{\nu} +
\A\, {x'}^{\mu}\, {x'}^{\nu} - \C\, ({\dot x}^{\mu}\, {x'}^{\nu}
+ {x'}^{\mu}\, {\dot x}^{\nu})\Big]\parallel.\nonumber \\
 &&{}
 \label{2.5}
 \eea
\medskip

\noindent The 4-vectors ${\dot x}^{\mu}(\tau ,\lambda )$ and
${x'}^{\mu}(\tau ,\lambda )$ are the null eigenvectors of the
Hessian matrix for every value of $\lambda$ except the end values
$\lambda = 0, \pi$ \footnote{The existence of the two null
eigenvalues, and of the first class constraints, is connected to the
$\tau , \lambda$ reparametrization invariance of the action; that is
the action is invariant under the following transformations
\begin{eqnarray*}
 \delta\tau &=& \tilde\tau (\tau ,\lambda ) - \tau,\qquad
  \delta\lambda = \tilde\lambda(\tau ,\lambda) - \lambda,\qquad
  \tilde\lambda (\tau ,\vec 0) =
0,\quad \tilde\lambda(\tau ,\pi ) = \pi,\nonumber \\
 \delta x^{\mu}(\tau ,\lambda ) &=& \tilde x^{\mu}(\tilde\tau ,\tilde\lambda ) -
x^{\mu}(\tau ,\lambda ) = \delta_0 x^{\mu}(\tau ,\lambda ) + {\dot
x}^{\mu}(\tau ,\lambda)\, \delta\tau + {x'}^{\mu}(\tau , \lambda )\,
\delta\lambda = 0,
 \end{eqnarray*}
\noindent where the last formula expresses the fact that
$x^{\mu}(\tau ,\lambda )$ is scalar under reparametrization.}. The
non-null eigenvalues are degenerate for $\lambda \not= 0, \pi$, and
are equal to $N\, x^{{'}2}(\tau ,\lambda )\over{\sqrt{- h(\tau
,\lambda )}}$. The non-null eigenvectors
$\zeta^{\mu}_{\epsilon}(\tau , \lambda )$, with $\epsilon = 1,2$,
are orthogonal to ${\dot x}^{\mu}$ and ${x'}^{\mu}$, i.e. to the
world-sheet, and so they are space-like.
\medskip

The variational principle for the action (\ref{2.1}), with the
variations $\delta_0 x^{\mu}(\tau ,\lambda )$ vanishing at $\tau =
\tau_1, \tau_2$, is

\beq
 \delta S = \int_{\tau_1}^{\tau_2}\, d\tau\, \int_0^{\pi}\,
 d\lambda\, L_{\mu}\, \delta_0 x^{\mu} - \int_{\tau_1}^{\tau_2}\,
d\tau\, \Pi_{\mu}\, \delta_0 x^{\mu} \vert_0^{\pi} = 0,
 \label{2.6}
 \eeq

\noindent and gives the following Euler-Lagrange equations and
boundary conditions ($\cir$ means evaluated on the solutions)

\bea
 L^{\mu}(\tau ,\lambda) &=& {\dot P}^{\mu} + {\Pi'}^{\mu} =
- W^{\mu\nu}\, \Big[{\ddot x}_{\nu} + \displaystyle{1\over{\B }}\,
\Big(\A\, {x''}_{\nu} - 2\, \C\, {{\dot x}'}_{\nu}\Big)\Big] =
\nonumber \\
 &=& N\, \partial_{\alpha}\, \Big(\sqrt{- h}\, h^{\alpha\beta}\, \partial_{\beta}
\, x^{\mu} \Big) \cir 0,\nonumber \\
 &&{}\nonumber \\
 \Pi_{\mu}\, \delta_0 x^{\mu}\, \vert_0^{\pi} &=& 0.
  \label{2.7}
 \eea

Only two of the equations (\ref{2.7}) are independent, for $\lambda
\not= 0,\pi$, since we have the Noether identities ${\dot x}_{\mu}\,
L^{\mu} \equiv 0$ and ${x'}_{\mu}\, L^{\mu} \equiv 0.$ The boundary
conditions in Eqs.(\ref{2.7}) have been studied in detail in
Refs.\cite{13a,14a}, where it is shown that in regular coordinates,
corresponding to a parametrization of the string world-sheet such
that the tangent vectors ${x\pri}\emu\varb$ and $\xpu\emu\varb$ do
not vanish and are independent, it amounts in requiring

\beq
 \Pi_{\mu}\, {x\pri}\emu\, \vert_0^{\pi} = N\, \sqrt{- h}\, \vert_0^{\pi} =
0.
 \label{2.8}
 \eeq
\medskip

As stressed by the authors of Refs.\cite{13a,14a}, the requirement
of regular coordinates is crucial for a consistent action principle.
The restriction from (\ref{2.10}) to (\ref{2.11}) is due to the
requirement that a variation of the boundaries must not violate the
condition $- h \geq 0$.
\medskip

Instead in the frequently used orthogonal gauge (OG) is defined by a
choice of parameters which satisfy, besides $- h(\tau ,\lambda) \geq
0$, the conditions

\bea
 \A + \B &=& \C = 0,\qquad
 \Rightarrow  h_{\alpha\beta} = \A\, \left(\begin{array}{cc}1&0\\
 0&-1\end{array}\right),\quad {\rm with}\quad \A \geq 0.\nonumber \\
 &&{}\nonumber \\
 &&\Downarrow\nonumber \\
 &&{}\nonumber \\
 L^{\mu} &=& N\, ({\ddot x}^{\mu} - {x''}^{\mu}) \cir 0,\qquad P^{\mu} =
 N\, {\dot x}^{\mu},\qquad \Pi^{\mu} = - N\, {x'}^{\mu},\nonumber \\
 W^{\mu\nu} &=& - \frac{N}{\A}\, \Big(\A\, \eta^{\mu\nu} - {\dot x}^{\mu}\, {\dot
x}^{\nu} + {x'}^{\mu}\, {x'}^{\nu}\Big).
 \label{2.9}
 \eea

At the end points the usual conditions of the OG gauge are

\beq
 {x\pri}\emu (\tau ,0)\, =\, {x\pri}\emu (\tau ,\pi)\, =\, 0,
 \qquad \Rightarrow\qquad \A(\tau ,0) = \A(\tau ,\pi) = 0,
 \label{2.10}
 \eeq

\noindent This in particular means that at the end points the
induced metric $h_{\alpha\beta}$ has zero rank.

\medskip

Let us remark that if one  chooses coordinates such that

\beq
 \A(\tau, \lambda)\, \geq\, 0,\qquad \B(\tau, \lambda)\, \leq\, 0,
 \label{2.11}
 \eeq

\noindent then the condition $h = 0$ implies two possible situations
at the end points:\medskip

i) $\B < 0$, $\A = 0$, $\C = 0$, with $\xpu\emu$ and ${x\pri}\emu$
independent and ${\xpu}\emu \not= 0$. This is a {\it regular case}
(the Jacobian of the map $\varb \rightarrow x\emu$ has maximal rank
2). In this case the rank of the induced metric $\parallel
h_{\alpha\beta}\parallel$ is 1, and the end points of the string
describe null surfaces \cite{13a}. There is the possible case
${\xpu}\emu =0$, which is a {\it singular case} (the Jacobian of the
map $\varb \rightarrow x\emu$ has rank 1).\medskip

ii) $\B = 0$, $\A = 0$, $\C = 0$, with ${x\pri}\emu$ collinear to
$\xpu\emu$. This is a {\it singular case} (where we may have
$\Pi\emu \not= 0$ as well as $\Pi\emu = 0$). The case ${x\pri}\emu =
0$, corresponding to the OG, may be considered as a particular case
of (ii): OG  is  a singular case.\medskip

As a consequence, to describe the solutions of the classical
equations of motion in a class of gauges including as a special case
the OG, we have to work with the class (ii), that is necessarily
with {\it singular coordinates}.\medskip

To check whether the boundary condition $h\, \vert_0^{\pi} = 0$ is
preserved in a singular case (since it was deduced in the regular
one), one may perform a transformation from regular coordinates to
those which will become singular at the end points, in the interior
of the interval $(0,\pi)$, that is from the class (i) to the class
(ii). As shown in Ref.\cite{14a}, and more explicitly in Section I1
of Appendix I of Ref.\cite{10a}, the Jacobian of the transformation
vanishes as $\sqrt{\lambda}$ in $\lambda = 0$ (and in an analogous
way in $\pm\pi$), so ensuring, {\sl a fortiori}, the vanishing of
the new $h$ at the end points.\medskip

So we will assume the boundary conditions (\ref{2.11}), with a
choice of coordinates falling into  class (ii). In order to
completely define the physical hypotheses, we will assume that the
total momentum of the string $P\emu$ be such that $P^2 \geq 0$, with
$P\emu \not= 0$. As shown in Section I2 Appendix I of
Ref.\cite{10a}, this ensures a unique solution at the end points of
the string with $P\emu$ and $\Pi\emu$ finite.\medskip

The previous discussion and the boundary conditions (\ref{2.8})
suggest the usefulness of the following extension from the interval
$(0,\pi)$ to $(-\pi,\pi)$

\beq
 x^{\mu}(\tau ,\lambda) = x^{\mu}(\tau , - \lambda ),
 \label{2.12}
 \eeq

\noindent and to the real line with $2\, \pi$ periodicity. Let us
stress that, with this kind of boundary conditions, the function
${x\pri}\emu\varb$, extended to all the real axis, may be
discontinuous in $\lambda = 0, \pi$.\medskip

The solutions of Eqs. (\ref{2.9}) satisfying Eqs.(\ref{2.10}) are

\bea
 x^{\mu}(\tau ,\lambda) &=& q^{\mu} +
 \displaystyle{P^{\mu}\over{\pi\,
N}}\, \tau + f^{\mu}(\tau + \lambda) + f^{\mu}(\tau - \lambda) =
\nonumber \\
 &=& q^{\mu} + \displaystyle{P^{\mu}\over{\pi\, N}}\, \tau +
\displaystyle{i\over{\sqrt{\pi\, N}}}\, \sum_{n\not= 0}\,
{\alpha}_n^{\mu}\, \exp{(- i\, n\, \tau)}\, \cos{n\lambda} =
\nonumber \\
 &=& \displaystyle{1\over{2}}\, \Big[Q^{\mu}(\tau + \lambda) + Q^{\mu}(\tau -
\lambda)\Big], \nonumber \\
&&{}\nonumber \\
 with && \left(\displaystyle{P^{\mu}\over{2\pi N}} +
\displaystyle{df^{\mu}(u)\over{du}}\right)^2 = 0,\qquad\qquad
f^{\mu}(u)= f^{\mu}(u+2n\pi),
 \label{2.13}
 \eea

\noindent where $u = \tau \pm \lambda,$ and where $P^{\mu} =
\int_0^{\pi} d\sigma P^{\mu}(\tau ,\lambda)$ is the conserved total
momentum. The last of Eqs. (\ref{2.13}) is a consequence of the OG
conditions, and

\bea
 Q^{\mu}(\tau) &=& x^{\mu}(\tau ,0) = q^{\mu} +
 \displaystyle{P^{\mu}\over{\pi\, N}}\, \tau + 2\,
 f^{\mu}(\tau),\qquad
 Q^{\mu}(\tau + 2\, \pi) = Q^{\mu}(\tau) +
 2\, \displaystyle{P^{\mu}\over{N}},\nonumber \\
 &&{}
 \label{2.14}
 \eea

\noindent is the coordinate of the end point at $\lambda = 0$. In
terms of $Q^{\mu}$ Eqs.(\ref{2.13}) imply
$\displaystyle{1\over{4}}\, {\dot Q}^2(\tau) = 0.$ The coordinates
of the other end point are

\beq
 x^{\mu}(\tau ,\pi) = x^{\mu}(\tau + \pi ,0) -
\displaystyle{P^{\mu}\over{N}} = Q^{\mu}(\tau + \pi) -
\displaystyle{P^{\mu}\over{N}}.
 \label{2.15}
 \eeq

\noindent In Ref.\cite{15a} it is shown that the end points suffer a
constant translation of ${{2\, P^{\mu}}\over{N}}$ for every
$\Delta\tau = 2\, \pi$, and that for $\Delta\tau = \pi$ the distance
between them is $P^{\mu}\over{N}$. Their motion is given by a double
helix with these periods. $Q^{\mu}(\tau)$ is a relevant function,
because the transverse conformal invariant oscillators defined in
Ref. \cite{16a} ${\bf A}_n = \sqrt{\displaystyle{N\over{2\, \pi}}}\,
\int_{-\pi}^{\pi}\, d\rho\, \displaystyle{d{\bf
Q}(\rho)\over{d\rho}}\, \exp{(i\, \pi\, n\displaystyle{Q^{+}
(\rho)\over{P^{+}}})}$, and the vertex of the dual models
$\exp{(iQ^{+}(z))}$ are defined in terms of it \cite{17a}. The
Cauchy problem for the equations (\ref{2.9}) is defined in
Ref.\cite{18a}.\medskip

The {\it residual invariance group} in the OG is given by the
conformal transformations holding the end points $\lambda = 0,\pi$
fixed:

\bea
 \tilde\tau &=& \tau_1 + \tau + g(\tau +\lambda) + g(\tau -\lambda) =
\tau_1 + \tau + \sum_{n\not= 0}\, a_n\, \cos{n\lambda}\, \exp{(-
i\, n\, \tau)},\nonumber \\
 &&{}\nonumber \\
 \tilde\lambda &=& \lambda + g(\tau +\lambda) - g(\tau -\lambda) = \lambda -
i\, \sum_{n\not= 0}\, a_n\, \sin{n\lambda}\, \exp{(- i\, n\, \tau)},
 \label{2.16}
 \eea

\noindent and the Jacobian is

\beq
 J = \displaystyle{\partial (\tilde\tau,\tilde\lambda)
\over{\partial (\tau,\lambda)}} = [1 + 2\, g'(\tau +\lambda)][1 +
2\, g'(\tau -\lambda)] \not= 0,
 \label{2.17}
 \eeq

\noindent or $1 + 2\, \displaystyle{dg(u)\over{du}} \not= 0.$ These
transformations leave invariant in form the wave equation in
Eqs.(\ref{2.9}) and the conditions (\ref{2.10}).
\medskip

To completely fix the gauge one has to add a further condition, for
instance of the kind \cite{19a}

\beq
 t_{\mu}\, \Big[x^{\mu}(\tau ,\lambda) - q^{\mu} - \displaystyle{P^{\mu}\tau
\over{\pi\, N}}\Big] = 0,
 \label{2.18}
 \eeq

\noindent where $t_{\mu}$ is a constant vector. In the usual {\it
light-cone gauge} one has $t^{\mu} = (1;0,0,1),$ so that we have
[compare with Eq. (\ref{2.13})]

\beq
 x^{+}(\tau ,\lambda) - q^{+} - \displaystyle{P^{+}\, \tau\over{\pi\, N}}
= 0, \quad A^{+} = A^o + A^{3},\quad {\rm implying}\quad f^{+}(u) =
0.
 \label{2.19}
 \eeq

\medskip

In Ref.\cite{20a} it was noticed that, while in a time-like gauge,
$t^{\mu} = (1;{\bf 0})$ and $f^o(u) = 0$, every solution of Eq.
(\ref{2.9}) can be made to satisfy equation (\ref{2.18}) because
$P^{0} \not= 0$, in the light-cone gauge solutions exist which
cannot satisfy equation (\ref{2.18}). See the discussion in
Ref.\cite{10a} on how to recover these solutions in the Hamiltonian
approach by introducing the following Poisson structure.\medskip

As a consequence of the previous discussion, in order to define a
Poisson structure in the phase-space in Refs.\cite{8a,10a,11a} the
following extension of the coordinates outside the interval $(0,
\pi)$ was used

\bea
 x^{\mu}\varb &=& x^{\mu}(\tau ,- \lambda,) = x^{\mu}(\tau ,\lambda +
2n\pi),\nonumber \\
 P^{\mu}\varb &=& P^{\mu}(\tau , - \lambda ) =
P^{\mu}(\tau ,\lambda + 2n\pi ),
 \label{2.20}
 \eea

\noindent where n is an integer, with the points $\lambda = 2\, n\,
\pi$, for any n (or $\lambda = (2\, n + 1)\, \pi$, for any n),
corresponding to the boundary values $x^{\mu}(\tau ,0)$ and
$P^{\mu}(\tau ,0)$ (or $x^{\mu}(\tau ,\pi)$ and $P^{\mu}(\tau ,
\pi)$) as limit values from the open interval $(0, \pi)$.\medskip

With the kind of chosen boundary conditions  and with this kind of
extension to the real axis, the functions ${x\pri}\emu\varb$ and
${P\pri}\emu\varb$ will have in $\lambda_{i}$ $(i=1,2; \lambda_1 =
0, \lambda_2 = \pi)$ a jump, not present in the OG. This means that
we must define the physical values of these functions as limit value
from the open interval $(0,\pi)$, since the Fourier series will
converge point-wise to the mean value of the left and right limits
at the end points, that is to unphysical values.

\bigskip

Following Ref.\cite{19a} we introduce the following Poisson
structure

\beq
 \{ x^{\mu}\varb, P^{\nu}(\tau ,\lambda') \} =
- \eta^{\mu\nu}\, \Delta_+(\lambda, \lambda')\quad \longrightarrow
\quad - \eta^{\mu\nu}\, \delta( \lambda - \lambda'),\quad {\rm for }
\ \lambda, \lambda' \in (0, \pi),
 \label{2.21}
 \eeq

\noindent where $\Delta_+(\lambda, \lambda')$ is an even delta
function with period $2 \pi$, defined in Eqs.(\ref{a1}) and
(\ref{a3}) of Appendix A.

\medskip

This definition implies a suitable definition of the functional
derivative

\beq
 \frac{\delta x_{\mu}(\lambda)}{\delta x\enu(\lambda')}
= \frac{\delta P_{\mu}(\lambda)}{\delta P\enu(\lambda')} =
\eta_{\mu\nu}\, \Delta_{+}(\lambda, \lambda').
 \label{2.22}
  \eeq
\medskip

More generally  the Poisson bracket for two canonical variables
$A(\lambda)$, $B(\lambda)$ has the form

\beq
 \{ A(\lambda), B(\lambda\pri) \} = \int_{0}^{\pi}\,
d\bar\lambda\, \Big[ \frac{\delta A(\lambda)}{\delta
P_{\mu}(\bar\lambda)}\,\, \frac{\delta B(\lambda\pri)}{\delta
x\emu(\bar\lambda)} - \frac{\delta A(\lambda)}{\delta
x\emu(\bar\lambda)}\,\, \frac{\delta B(\lambda\pri)}{\delta
P_{\mu}(\bar\lambda)}\Big].
 \label{2.23}
 \eeq

Since there can be a dependence on $x\pri (\tau ,\lambda)$, we must
check if some boundary term is present \cite{21a,22a}.

\medskip

The naive center-of-mass coordinates of the string are

\bea
 X^{\mu}(\tau) &=&
\displaystyle{1\over{2\, \pi}}\, \int_{-\pi}^{\pi}\, d\lambda\,
x^{\mu}\varb,\qquad
 P^{\mu} = \displaystyle{1\over{2}}\, \int_{-\pi}^{\pi}\,
 d\lambda\, P^{\mu}\varb,\nonumber \\
 &&{}
 \label{2.24}
 \eea

\noindent  where $P^{\mu}$ is the conserved generator of the
space-time translations. The following relative coordinates have
been introduced in Ref.\cite{10a}

\bea
 y^{\mu}\varb &=& - {x'}^{\mu}\varb = - y^{\mu}(\tau ,- \lambda
 ),\nonumber \\
{\cal P}^{\mu}\varb &=& \int_0^{\lambda}\, d\lambda'\, P^{\mu}(\tau
, \lambda') - \displaystyle{\lambda\over{\pi}}\, P^{\mu} = - {\cal
P}^{\mu}(\tau ,- \lambda ) = {\cal P}\emu(\tau ,\lambda + 2\, n\,
\pi ),
 \label{2.25}
 \eea

\noindent with the following properties

\bea
 &&\int_{-\pi}^{\pi}\, d\lambda\,  y^{\mu}\varb = \int_{-\pi}^{\pi}\,
 d\lambda\, {\cal P}^{\mu}\varb = 0,\nonumber \\
  &&{\cal P}\emu (0) = {\cal P}^{\mu}(\pm\pi) = 0,\quad \rightarrow
  \int_{-\pi}^{\pi}\, d\lambda\, {{\cal P}'}^{\mu}\varb = 0,
  \label{2.26}
  \eea

\noindent and the following inversion

\bea
  x^{\mu}\varb &=& X^{\mu}(\tau) + \displaystyle{1\over{2\,
  \pi}}\, \int_{-\pi}^{\pi}\, d\lambda_1\, \int_{0}^{\lambda_{1}}\,
  d\lambda_2\, y^{\mu}(\tau ,\lambda_2) - \int_0^{\lambda}\,
  d\lambda_{2}\, y^{\mu}(\tau ,\lambda_2),\nonumber \\
   P^{\mu}\varb &=& \displaystyle{1\over{\pi}}\, P^{\mu} + {{\cal P}'}^{\mu}\varb.
 \label{2.27}
 \eea
\medskip

It may be checked that the coordinates (\ref{2.24}) and (\ref{2.25})
constitute a basis of canonical variables

\beq
 \{ X^{\mu}, P^{\nu} \} = - \eta^{\mu\nu},\qquad
\{ y^{\mu}\varb, {\cal P}^{\nu}(\tau ,\lambda')\} = -
\eta^{\mu\nu}\, \Delta_- (\lambda,\lambda'),
 \label{2.28}
 \eeq

\noindent with all the other Poisson brackets vanishing. Here
$\Delta_- (\lambda,\lambda')$ is the odd delta function with period
$2 \pi$ defined in Eq.(\ref{a2}).
\medskip

The constraints implied by equation (\ref{2.6}) are

\bea
 {\bar \chi}_1\varb &=& {\bar \chi}_1(\tau ,- \lambda ) = P^2\varb + N^2\, {x'}^2\varb
\approx 0,\nonumber \\
 {\bar \chi}_2\varb &=& - {\bar \chi}_2(\tau ,- \lambda ) =
P\varb \cdot {x'}\varb \approx 0.
 \label{2.29}
 \eea

\noindent The following equivalent expression is given in terms of
the functions $A^{\mu}_{\pm}(\tau, \lambda)\, =\, A^{\mu}_{\mp}(\tau
,-\lambda )\, =\, P^{\mu}\varb \pm  N\, {x'}^{\mu}\varb\, =
\displaystyle{1\over{\pi}}\, P^{\mu} + {{\cal P}'}^{\mu}\varb \mp
N\, y^{\mu}\varb\, =\, \displaystyle{\partial\
\over{\partial\lambda}}\, B^{\mu}_{\pm}\varb$ \footnote{As shown in
Ref.\cite{10a} one has $x^{\mu}(\tau ,\lambda )\, =\, {1\over {2\,
N}}\, \Big( B^{\mu}_{+}(\tau ,\lambda ) - B^{\mu}_{-}(\tau ,\lambda
) \Big)$ and $P^{\mu}(\tau ,\vec \lambda )\, =\, {1\over 2}\, \Big(
A^{\mu}_{+}(\tau ,\lambda ) + A^{\mu}_{-}(\tau ,\lambda )\Big)$ with
$B^{\mu}_{\pm}\varb\, =\, - B^{\mu}_{\mp}(\tau ,- \lambda )\, =\,
\displaystyle{\lambda\over{\pi}}\, P^{\mu} + {\cal P}^{\mu}\varb \pm
N\, x^{\mu}\varb$, $B^{\mu}_{\pm}(\tau ,\lambda + 2\, n\, \pi )\,
=\, B^{\mu}_{\pm}\varb + 2\, n\, P^{\mu}$, $B^{\mu}_{\pm}(\tau ,\pi
) - B^{\mu}_{\pm}(\tau ,- \pi )\, =\, \int_{-\pi}^{\pi}\, d\lambda\,
A^{\mu}_{\pm}\varb = 2\, P^{\mu}$. The Poisson brackets of the
functions $A^{\mu}_{\pm}$ are $\{ A^{\mu}_{\pm}(\tau ,\lambda_1),
A^{\nu}_{\pm}(\tau ,\lambda_2) \}\, =\, \mp N\, \eta^{\mu\nu}\,
\Big( {\Delta'}_+(\lambda_1, \lambda_2) + {\Delta'}_-(\lambda_1,
\lambda_2)\Big)$, $\{ A^{\mu}_{\pm}(\tau ,\lambda_1),
A^{\nu}_{\mp}(\tau ,\lambda_2) \}\, =\, \mp N\, \eta^{\mu\nu}\,
\Big( {\Delta'}_+(\lambda_1, \lambda_2) - {\Delta'}_-(\lambda_1,
\lambda_2) \Big)$.}

\beq
 {\bar \chi}_{\pm}\varb = {\bar \chi}_{\mp}(\tau ,- \lambda ) =
 {A^2}_{\pm}\varb \approx 0,\qquad {\bar \chi}_1 = \displaystyle{1\over{2}}\,
 ( {\bar \chi}_+ + {\bar \chi}_- ),\qquad {\bar \chi}_2 = \displaystyle{1\over{4\, N}}\, (
{\bar \chi}_+ - {\bar \chi}_- ).
 \label{2.30}
 \eeq

They satisfy the algebra \cite{10a}

\bea
  \{ {\bar \chi}_{\pm}(\tau ,\lambda_1),
{\bar \chi}_{\pm}(\tau ,\lambda_2) \} &=& \mp 2\, N\, \Big({\bar
\chi}_{\pm}(\tau ,\lambda_1) + {\bar \chi}_{\pm}(\tau , \lambda_2)
\Big) \,\, \Big({\Delta'}_+(\lambda_1, \lambda_2) +
{\Delta'}_-(\lambda_1, \lambda_2) \Big), \nonumber \\
 \{ {\bar \chi}_+(\tau ,\lambda_1), {\bar \chi}_-(\tau ,\lambda_2) \}
 &=& - 2\, N\, \Big( {\bar \chi}_+(\tau ,\lambda_1) + {\bar \chi}_-(\tau ,\lambda_2)
 \Big)\,\, \Big( {\Delta'}_+(\lambda_1, \lambda_2) -
{\Delta'}_-(\lambda_1, \lambda_2) \Big),\nonumber \\
 &&{}\nonumber \\
  \{ {\bar \chi}_2(\tau ,\lambda_1), {\bar \chi}_2(\tau ,\lambda_2) \} &=&
 - \Big({\bar \chi}_2(\tau ,\lambda_2)\, {\Delta'}_+(\lambda_1, \lambda_2) +
 {\bar \chi}_2(\tau ,\lambda_1)\, {\Delta'}_{-}(\lambda_1, \lambda_2)\Big).\nonumber \\
 &&{}
 \label{2.31}
 \eea

Therefore the constraints are $1^{st}$-class, but they are in weak
involution; the algebra (\ref{2.31}) is the universal Dirac algebra
of reparametrization.

\bigskip

The Poincare' generators are [$Y^{\mu}(\tau ,\lambda ) =
x^{\mu}(\tau ,\lambda ) - X^{\mu}(\tau )$]

\bea
 P^{\mu} &=& {1\over 2}\, \int^{\pi}_{-\pi} d\lambda\, P^{\mu}(\tau
 ,\lambda ),\nonumber \\
 J^{\mu\nu} &=& {1\over 2}\, \int^{\pi}_{-\pi} d\lambda\, [x^{\mu}(\tau ,\lambda )\,
 P^{\nu}(\tau ,\vec \lambda ) - x^{\nu}(\tau ,\lambda )\, P^{\mu}(\tau
 ,\lambda )] = L^{\mu\nu} + S^{\mu\nu} =\nonumber \\
 &=& X^{\mu}(\tau )\, P^{\nu} - X^{\nu}(\tau )\, P^{\mu} + {1\over
 2}\, \int^{\pi}_{-\pi} d\lambda\, [Y^{\mu}(\tau ,\lambda )\, {\cal P}^{\nu}(\tau
 ,\lambda ) - Y^{\nu}(\tau ,\lambda )\, {\cal P}^{\mu}(\tau ,\lambda )].
 \label{2.32}
 \eea
\bigskip

After this review of the standard descriptions of the open Nambu
string, we will reformulate it as a parametrized Minkowski theory in
the next Section. We will define different configuration variables,
but we will save the periodicity conditions and the Poisson
structure with the $\Delta_{\pm}$ delta functions.

\vfill\eject

\section{The Nambu String as a Parametrized Minkowski Theory.}

As  was done for scalar relativistic particles in
Refs.\cite{1a,3a,4a,5a,6a}, we can also describe the open Nambu
string by means of a parametrized Minkowski theory.\medskip

To formulate this theory we need the {\it 3+1 point of view}, in
which we assign: a) the world-line of an arbitrary time-like
observer; b) an admissible 3+1 splitting of Minkowski space-time,
namely a nice foliation with space-like instantaneous 3-spaces (i.e.
a clock synchronization convention). This allows one to define a
{\it global non-inertial frame} centered on the observer and to use
observer-dependent Lorentz-scalar {\it radar 4-coordinates}
$\sigma^A = (\tau ;\sigma^r)$, where $\tau$ is a monotonically
increasing function of the proper time of the observer and
$\sigma^r$ are curvilinear 3-coordinates on the instantaneous
3-spaces $\Sigma_{\tau}$ having the observer as origin. If $x^{\mu}
\mapsto \sigma^A(x)$ is the coordinate transformation from the
inertial Cartesian 4-coordinates $x^{\mu}$ to radar coordinates, its
inverse $\sigma^A \mapsto x^{\mu} = z^{\mu}(\tau ,\sigma^r)$ defines
the {\it embedding} functions $z^{\mu}(\tau ,\sigma^r)$ describing
the 3-spaces $\Sigma_{\tau}$ as embedded 3-manifold into Minkowski
space-time. The induced 4-metric on $\Sigma_{\tau}$ is the following
functional of the embedding ${}^4g_{AB}(\tau ,\sigma^r) =
[z^{\mu}_A\, \eta_{\mu\nu}\, z^{\nu}_B](\tau ,\sigma^r)$, where
$z^{\mu}_A = \partial\, z^{\mu}/\partial\, \sigma^A$. While the
4-vectors $z^{\mu}_r(\tau ,\sigma^u)$ are tangent to
$\Sigma_{\tau}$, so that the unit normal $l^{\mu}(\tau ,\sigma^u)$
is proportional to $\epsilon^{\mu}{}_{\alpha \beta\gamma}\,
[z^{\alpha}_1\, z^{\beta}_2\, z^{\gamma}_3](\tau ,\sigma^u)$, we
have $z^{\mu}_{\tau}(\tau ,\sigma^r) = [N\, l^{\mu} + N^r\,
z^{\mu}_r](\tau ,\sigma^r)$ ($N(\tau ,\sigma^r) = \sgn\,
[z^{\mu}_{\tau}\, l_{\mu}](\tau ,\sigma^r)$ and $N_r(\tau ,\sigma^r)
= - \sgn\, g_{\tau r}(\tau ,\sigma^r)$ are the lapse and shift
functions).\medskip

The foliation is nice and admissible if it satisfies the conditions:
\hfill\break
 1) $N(\tau ,\sigma^r) > 0$ in every point of
$\Sigma_{\tau}$ (the 3-spaces never intersect, avoiding the
coordinate singularity of Fermi coordinates);\hfill\break
 2) ${}^4g_{\tau\tau}(\tau ,\sigma^r) > 0$, so as to avoid the
 coordinate singularity of the rotating disk, and with the positive-definite 3-metric
${}^3g_{rs}(\tau ,\sigma^u) = -  {}^4g_{rs}(\tau ,\sigma^u)$ having
three positive eigenvalues (these are the M$\o$ller conditions
\cite{3a});\hfill\break
 3) all the 3-spaces $\Sigma_{\tau}$ must tend to the same space-like
 hyper-plane at spatial infinity (so that there are always asymptotic inertial
observers to be identified with the fixed stars).\medskip

In parametrized Minkowski theories one considers any isolated system
(particles, strings, fields, fluids) admitting a Lagrangian
description, because it allows, through the coupling to an external
gravitational field, the determination of the matter energy-momentum
tensor and of the ten conserved Poincare' generators $P^{\mu}$ and
$J^{\mu\nu}$ (assumed finite) of every configuration of the system.
Then one replaces the external gravitational 4-metric in the coupled
Lagrangian with the 4-metric $g_{AB}(\tau ,\sigma^r)$ of an
admissible 3+1 splitting of Minkowski space-time and  the matter
fields with new ones knowing the instantaneous 3-spaces
$\Sigma_{\tau}$. For instance for a relativistic particle with
world-line $x^{\mu}(\tau )$ we must make a choice of its energy
sign: then it will be described by 3-coordinates $\eta^r(\tau )$
defined by the intersection of the world-line with $\Sigma_{\tau}$:
$x^{\mu}(\tau ) = z^{\mu}(\tau ,\eta^r(\tau ))$. Differently from
all the previous approaches to relativistic mechanics, the dynamical
configuration variables are the 3-coordinates $\eta^r_i(\tau)$ and
not the world-lines $x^{\mu}_i(\tau)$ (to rebuild them in an
arbitrary frame we need the embedding defining that frame!).\medskip

With this procedure we get a Lagrangian depending on the given
matter and on the embedding $z^{\mu}(\tau ,\sigma^r)$, which is
invariant under frame-preserving diffeomorphisms. As a consequence,
there are four first-class constraints (an analogue of the
super-Hamiltonian and super-momentum constraints of canonical
gravity) implying that the embeddings $z^{\mu}(\tau ,\sigma^r)$ are
{\it gauge variables}, so that all the admissible non-inertial or
inertial frames are gauge equivalent, namely physics does {\it not}
depend on the clock synchronization convention and on the choice of
the 3-coordinates $\sigma^r$.

\medskip

Let us now consider the Nambu action re-written on space-like
hyper-surfaces $\Sigma_{\tau}$, leaves of an admissible 3+1
splitting of Minkowski space-time, identified by the embeddings
$x^{\mu} = z^{\mu}(\tau ,\vec \sigma )$. Let the world-sheet of the
string in Minkowski space-time be parametrized as

\bea
 x^{\mu}(\tau ,\lambda ) &=& z^{\mu}(\tau , \eta^r(\tau , \lambda
 )),\nonumber \\
 &&{}\nonumber \\
 &&x^{{'} \mu}(\tau, \lambda) = z^{\mu}_r(\tau , \eta^u(\tau , \lambda
 ))\, \eta^{{'} r}(\tau, \lambda),\nonumber \\
 &&{\dot x}^{\mu}(\tau, \lambda) = z^{\mu}_{\tau}(\tau , \eta^u(\tau , \lambda
 )) + z^{\mu}_r(\tau , \eta^u(\tau , \lambda
 ))\, {\dot \eta}^r(\tau, \lambda),\nonumber \\
 &&{}\nonumber \\
 &&{\dot x}^2(\tau, \lambda) = {}^4g_{\tau\tau}(\tau, \eta^u(\tau, \lambda))
 + 2\, {}^4g_{\tau r}(\tau, \eta^u(\tau, \lambda))\, {\dot \eta}^r(\tau, \lambda)
 +\nonumber \\
 &+& {}^4g_{rs}(\tau, \eta^u(\tau, \lambda))\, {\dot \eta}^r(\tau, \lambda)\,
 {\dot \eta}^s(\tau, \lambda),\nonumber \\
 &&x^{{'} 2}(\tau, \lambda) = {}^4g_{rs}(\tau, \eta^u(\tau, \lambda)\,
 \eta^{{'} r}(\tau, \lambda))\, \eta^{{'} s}(\tau, \lambda),\nonumber \\
 &&{\dot x}(\tau, \lambda) \cdot x^{'}(\tau, \lambda) = {}^4g_{\tau
 s}(\tau, \eta^u(\tau, \lambda)) + {}^4g_{rs}(\tau, \eta^u(\tau, \lambda))\,
 {\cdot \eta}^r(\tau, \lambda))\, \eta^{{'} s}(\tau,
 \lambda).\nonumber \\
 &&{}
 \label{3.1}
 \eea

Namely on each instantaneous 3-space $\Sigma_{\tau}$ the string is
described by the curvilinear coordinates $\eta^r(\tau , \lambda )$
with $\lambda \in (0, \pi )$ and by conjugate 3-momenta
$\kappa_r(\tau, \lambda)$. We use the notation ${\dot \eta}^r(\tau
,\lambda ) = {{\partial \eta^r(\tau ,\lambda )}\over {\partial
\tau}}$ and $\eta^{{'}\, r}(\tau ,\lambda ) = {{\partial \eta^r(\tau
,\lambda )}\over {\partial \lambda}}$ for the partial derivatives of
$\eta^r(\tau , \lambda )$ with respect to $\tau$ and $\lambda$,
respectively. We have two different theories for the two signs of
the total energy $sign\, P^o$. We shall consider only the sector
with $P^o > 0$.
\medskip

Like the world-lines in the case of particles, the coordinates
$x^{\mu}(\tau, \lambda)$ are derived quantities and the old
4-momenta $P^{\mu}(\tau, \lambda)$ are not defined. In the case of
particles one can define a derived quantity $P^{\mu}$ satisfying the
mass-shell constraint $P^2 = m^2$. Also in the string case in the
inertial frames one can define the following 4-vector

\bea
 P^{\mu}(\tau, \lambda) &=& \pm\, l^{\mu}(\tau, \eta^u(\tau,
 \lambda))\, \sqrt{ N^2\, {}^4g_{rs}(\tau, \eta^u(\tau, \lambda))\,
 \eta^{{'}r}(\tau, \lambda)\, \eta^{{'} s}(\tau, \lambda) -
 \gamma^{rs}(\tau, \eta^u(\tau, \lambda))\, \kappa_r(\tau, \lambda)\,
 \kappa_s(\tau, \lambda)} + \nonumber \\
 &+& \kappa_r(\tau, \lambda)\, \gamma^{rs}(\tau, \eta^u(\tau,
 \lambda))\, z^{\mu}_s(\tau, \eta^u(\tau, \lambda)),
 \label{3.1a}
 \eea

\noindent satisfying $P^2(\tau, \lambda) + N^2\, x^{{'} 2}(\tau,
\lambda) = 0$.

\medskip

Only the transversality constraints, i.e. the counterpart of
$P(\tau,\lambda ) \cdot x^{'}(\tau ,\lambda ) \approx 0$, will
remain. They are the generators of the passive diffeomorphisms along
the string and say that the longitudinal degree of freedom is a
gauge variable.

\medskip
The action (\ref{2.1}) becomes

\bea
 S &=& \, \int d\tau\, \int_0^{\pi}\, d\lambda\, L(\tau ,\lambda
 ),\nonumber \\
 &&{}\nonumber \\
 L(\tau , \lambda  ) &=& -    N\, \int d^3\sigma\, \delta^3(
 \sigma^a - \eta^a(\tau ,\lambda ))\,
 \Big[ \Big(- {}^4g_{\tau\tau}\, {}^4g_{rs} +
 {}^4g_{\tau r}\, {}^4g_{\tau s}\Big)\, \eta ^{\prime r}(\tau ,\lambda
 )\, \eta ^{\prime s}(\tau ,\lambda ) +\nonumber \\
 &+& \Big( {}^4g_{rs}\, {}^4g_{uv} - {}^4g_{ru}\, {}^4g_{sv}\Big)\, \dot \eta ^{r}(\tau ,\lambda )\,
 \eta ^{\prime s}(\tau ,\lambda )\, \dot \eta^{u}(\tau ,\lambda )\,
 \eta ^{\prime v}(\tau ,\lambda ) +\nonumber \\
 &+& 2\, \Big( {}^4g_{rs}\, {}^4g_{\tau u} - {}^4g_{us}\, {}^4g_{\tau r}\Big)\,
 \dot \eta ^{r}(\tau ,\lambda )\, \eta ^{\prime s}(\tau ,\lambda
 )\,\eta ^{\prime u}(\tau ,\lambda ) \Big]
 ^{\frac 12}(\tau , \sigma^u ) =\nonumber \\
 &{\buildrel {def}\over =}& - N\, \int d^3\sigma\, \delta^3(\sigma^a - \eta^a(\tau
 ,\lambda ))\, \sqrt{-G(z^{\rho}(\tau, \sigma^u), \eta^u(\tau
 ,\lambda ))}.
 \label{3.2}
 \eea

\medskip

The canonical momenta conjugate to the configuration variables
$z^{\mu}(\tau ,\vec \sigma )$ and $\vec \eta (\tau ,\lambda )$ are
(${\cal N}(\tau ,\vec \sigma ) = \sqrt{{{g}\over {\gamma}}}(\tau
,\vec \sigma )$, ${\cal N}^{r}(\tau ,\vec \sigma ) =
\gamma^{rs}(\tau ,\vec \sigma )\, {}^4g_{\tau s}(\tau ,\vec \sigma
)$ are the lapse and shift functions respectively; $\gamma^{rs}$ is
the inverse of ${}^4g_{rs}$)

\bea
 \rho_{\mu}(\tau , \sigma^u ) &=& - {{\delta S}\over {\delta\,
 z^{\mu}_{\tau}(\tau , \sigma^u )}} =\nonumber \\
 &=& \int_0^{\pi} d\lambda\, \delta^3(\sigma^a - \eta^a(\tau ,\lambda ))\,
 {N\over {\sqrt{-G(z^{\rho}(\tau, \sigma^u), \eta^u(\tau
 ,\lambda ))}}}\nonumber \\
 &&\times  \Big( l_{\mu}(\tau , \sigma^u )\, \Big[ -
 {\cal N}\, {}^4g_{rs}\, \eta^{{'}\, r}(\tau
 ,\lambda )\, \eta^{{'}\, s}(\tau ,\lambda )\Big] (\tau
 ,\vec \sigma ) +\nonumber \\
 &+& z_{s \mu}(\tau , \sigma^u )\, \gamma^{s
 r}(\tau , \sigma^u )\, \Big[ - {}^4g_{ru}\, ({\cal
 N}^{u} + {\dot \eta}^{u}(\tau ,\lambda ))\,
 {}^4g_{mn}\, \eta^{{'} m}(\tau ,\lambda )\,
 \eta^{{'} n}(\tau ,\lambda) +\nonumber \\
 &+& {}^4g_{ru}\, \eta^{{'} u}(\tau ,\lambda )\, {}^4g_{mn}\,
 ({\cal N}^{m} + {\dot \eta}^{m}(\tau ,\lambda ))\,
 \eta^{{'} n}(\tau ,\lambda )\Big] (\tau , \sigma^u
 )\Big),\nonumber \\
 &&{}\nonumber \\
 \kappa_{r}(\tau ,\lambda ) &=& - {{\delta S}\over
 {\delta\, {\dot \eta}^r(\tau ,\lambda )}} =\nonumber \\
 &=&{N\over {\sqrt{-G(z^{\rho}(\tau, \eta^u(\tau ,\lambda )), \eta^u(\tau
 ,\lambda ))}}}\, \Big[ ({}^4g_{rs}\, {}^4g_{u
 v} - {}^4g_{ru}\, {}^4g_{sv})\, {\dot \eta}^{u}(\tau ,\lambda ) + \nonumber \\
 &+& {}^4g_{rs}\, {}^4g_{\tau v} - {}^4g_{\tau r}\, {}^4g_{sv}\Big] (\tau ,
 \eta^u(\tau ,\lambda ))\, \eta^{{'} s}(\tau
 ,\lambda )\, \eta^{{'} v}(\tau ,\lambda ),
 \label{3.3}
 \eea

\noindent with $\{ \eta^{r}(\tau ,\lambda ), \kappa_{ s}(\tau
,\lambda^{'}) \} = - \delta^r_s\, \delta (\lambda - \lambda^{'})$.
\medskip

With the extension $\eta^r(\tau ,\lambda ) = \eta^r(\tau ,-\lambda )
=  \eta^r(\tau ,\lambda + 2n\pi )$, $\kappa_r(\tau ,\lambda ) =
\kappa_r(\tau ,-\lambda ) = \kappa_r(\tau ,\lambda + 2n\pi )$, we
have $\{ \eta^r(\tau ,\lambda ), \kappa_s(\tau ,\lambda_1) \} =
\delta^r_s\, \Delta_+(\lambda ,\lambda_1) = \delta^r_s\, \delta
(\lambda - \lambda_1)$ for $\lambda ,\lambda_1 \in (0, \pi )$ with
the Poisson structure of Eq.(\ref{2.21}).
\medskip

Besides the constraints implying the independence from the foliation

\bea
 {\cal H}_{\mu}(\tau , \sigma^u ) &=& \rho_{\mu}(\tau
 , \sigma^u ) - \int_0^{\pi} d\lambda\, \delta^3(\sigma^a -
 \eta^a(\tau ,\lambda ))\, \Big( z_{s \mu}(\tau , \sigma^u )\,
 \gamma^{sr}(\tau , \sigma^u )\, \kappa_r(\tau ,\lambda ) +\nonumber \\
 &+& l_{\mu}(\tau , \sigma^u )\, \sqrt{N^2\, {}^4g_{rs}(\tau
 , \sigma^u )\, \eta^{{'} r}(\tau ,\lambda )\, \eta^{{'} s}(\tau
 ,\lambda ) - \gamma^{rs}(\tau , \sigma^u )\,
 \kappa_r(\tau ,\lambda )\, \kappa_s(\tau ,\lambda )
 }\Big) \approx 0,\nonumber \\
 &&{}
 \label{3.4}
 \eea

\noindent we get the following transversality constraints

\beq
 \bar \chi (\tau ,\lambda ) = - \kappa_r(\tau ,\lambda )
 \eta^{{'} r}(\tau ,\lambda )   \approx 0.
 \label{3.5}
 \eeq

All the constraints are first class. In particular we have the
following Poisson brackets implying the diffeomorphism algebra (the
analogue of the last line in Eqs.(\ref{2.31}))

\bea
 \{ \bar \chi (\tau ,\lambda ), \bar \chi (\tau , \lambda^{'}) \}
 &=& \bar \chi (\tau ,\lambda )\, {{\partial\, \delta (\lambda -
 \lambda^{'})}\over {\partial \lambda}} - \bar \chi (\tau
 ,\lambda^{'})\,  {{\partial\, \delta (\lambda -
 \lambda^{'})}\over {\partial \lambda^{'}}} =\nonumber \\
 &=& \bar \chi (\tau ,\lambda )\, \Delta^{'}_-(\lambda ,\lambda')
 + \bar \chi (\tau ,\lambda' )\, \Delta^{'}_{+}(\lambda ,\lambda')
 \approx 0.
 \label{3.6}
 \eea

The Poincare' generators are

\bea
 P^{\mu} &=& \int d3\sigma\, \rho^{\mu}(\tau , \sigma^u
 ),\nonumber \\
 J^{\mu\nu} &=& \int d3\sigma\, \Big[z^{\mu}\, \rho^{\nu} -
 z^{\nu}\, \rho^{\mu}\Big](\tau , \sigma^u ).
 \label{3.7}
 \eea

The description of the string in non-inertial frames could be done
by using Ref.\cite{3a}. Instead in the next Section we will study
its rest-frame instant form.

\vfill\eject

\section{The Rest-Frame Instant Form of the Nambu String.}

If we restrict ourselves to inertial frames, we can define the {\it
inertial rest-frame instant form of dynamics for isolated systems}
\cite{4a,5a,6a} by choosing the 3+1 splitting corresponding to the
intrinsic inertial rest frame of the isolated system centered on an
inertial observer: the instantaneous 3-spaces, named Wigner 3-space
due to the fact that the 3-vectors inside them are Wigner spin-1
3-vectors, are orthogonal to the conserved 4-momentum $P^{\mu}$ of
the configuration.
\medskip

As said in the Introduction in the inertial rest frames we can give
the final solution to the old problem of the relativistic extension
of the Newtonian center of mass of an isolated system. In its rest
frame there are {\it only} three notions of collective variables,
which can be built by using {\it only} the Poincare' generators
(they are {\it non-local} quantities knowing the whole
$\Sigma_{\tau}$): the canonical non-covariant Newton-Wigner center
of mass (or center of spin), the non-canonical covariant
Fokker-Pryce center of inertia and the non-canonical non-covariant
M$\o$ller center of energy. All of them tend to the Newtonian center
of mass in the non-relativistic limit. As shown in Refs.\cite{4a}
these three variables can be expressed as known functions of the
rest time $\tau$, of the canonically conjugate Jacobi data (frozen
Cauchy data) $\vec z = M\, {\vec x}_{NW}(0)$ (${\vec x}_{NW}(\tau )$
is the standard Newton-Wigner 3-position) and $\vec h = \vec P/M$
($\{ z^i, h^j \} = \delta^{ij}$), of the invariant mass $M = \sqrt{
P^2}$ of the system and of its rest spin ${\vec {\bar S}}$.\medskip

As a consequence, every isolated system (i.e. a closed universe) can
be visualized as a decoupled non-covariant collective (non-local)
pseudo-particle described by the frozen Jacobi data $\vec z$, $\vec
h$ carrying a {\it pole-dipole structure}, namely the invariant mass
and the rest spin of the system, and with an associated {\it
external} realization of the Poincare' group. The universal breaking
of Lorentz covariance is connected to the decoupled non-local
collective variable $\vec z$. As already said in each Wigner 3-space
there is an {\it unfaithful inner} realization of the Poincare'
algebra, whose generators $M$, $\vec k \approx 0$, ${\vec {\bar
S}}$, ${\vec {\cal K}} \approx 0$ are built by using the
energy-momentum tensor of the isolated system. The three pairs of
second class constraints $\vec k \approx 0$, ${\vec {\cal K}}
\approx 0$ (the inner Lorentz boosts are interaction dependent),
eliminate the six degrees of freedom of the inner center of mass and
identify the observer origin of the 3-coordinates $\sigma^r$ with
the covariant non-canonical Fokker-Pryce center of inertia. The
invariant mass $M$, which depends only on the Wigner-covariant
relative variables (like the rest spin ${\vec {\bar S}}$), is the
effective Hamiltonian inside the Wigner 3-spaces as shown in
Ref.\cite{4a}.

\medskip

The Wigner hyper-planes are defined by the embeddings

\bea
 z_W^{\mu}(\tau ,\vec \sigma ) &=& Y^{\mu}(\tau ) +
\epsilon^{\mu}_r(P)\, \sigma^r,\nonumber \\
 &&{}\nonumber \\
 Y^{\mu}(\tau ) &=& u^{\mu}(P)\, \tau ,\qquad z^{\mu}_{W\tau}(\tau
 ,\vec \sigma ) = u^{\mu}(P),\,\, z^{\mu}_{Wr}(\tau ,\vec \sigma )
 = \epsilon^{\mu}_r(P),\nonumber \\
 &&{}\nonumber \\
 \epsilon _{o}^{\mu }(u(P)) &=&u^{\mu }(P)=P^{\mu}/\sqrt{P^{2}}
 = h^{\mu} = (\sqrt{1 + {\vec h}^2}; \vec h) =
 \epsilon^{\mu}_o(\vec h),\nonumber \\
 \epsilon _{r}^{\mu }(u(P)) &=& (-u_{r}(P);\delta _{r}^{i}-{\frac{{
 u^{i}(P)\,u_{r}(P)}}{{1+u^{o}(P)}}}) = \epsilon^{\mu}(\vec h),  \notag \\
 &&{}  \notag \\
 \epsilon _{\mu }^{o}(u(P)) &=&\eta ^{oB}\,\eta _{\mu \nu }\,\epsilon
 _{B}^{\nu }(u(P)) = u_{\mu }(P) = h_{\mu},\nonumber \\
 \epsilon_{\mu }^{r}(u(P)) &=& \eta ^{rB}\,\eta _{\mu \nu }\,\epsilon _{B}^{\nu
 }(u(P)) = \epsilon^r_{\mu}(\vec h).
 \label{4.1}
 \eea

\noindent where $Y^{\mu}(\tau )$ is the inertial observer
corresponding to the Fokker-Pryce  4-center of inertia of the
string. The space-like 4-vectors $\epsilon _{r}^{\mu }(\vec h)$
together with the time-like one $\epsilon _{o}^{\mu }(\vec h)$ are
the columns of the standard Wigner boost for time-like Poincare'
orbits \footnote{ It sends the time-like four-vector $P^{\mu }$ to
its rest-frame form $ \overset{\circ }{P}{}^{\mu }=\eta
\,\sqrt{P^{2}}(1;\vec{0})$, where $\eta =sign\,P^{o}$. From now on
we restrict ourselves to positive energies, i.e. $\eta =1$. While
$\epsilon _{o}^{\mu }(u(P))$ and $ \epsilon _{\mu }^{o}(u(P))$ are
4-vectors, $\epsilon _{r}^{\mu }(u(P))$ have more complex
transformation properties under Lorentz transformations.}.
\medskip

The open string and its end points have the following representation
[ ${\dot {\vec \eta}}_A(\tau ) = {{\partial}\over {\partial \tau}}\,
\vec \eta (\tau ,\lambda ){|}_{\lambda = A}$, $A= 0, \pi$]

\bea
 x^{\mu}(\tau ,\lambda ) &=& z_W^{\mu}(\tau , \vec \eta (\tau
 ,\lambda )) = Y^{\mu}(\tau ) + \epsilon^{\mu}_r(\vec h)\, \eta^r(\tau
 ,\lambda ),\nonumber \\
  &&{}\nonumber \\
 Y^{\mu}(\tau ) &=& \Big(\sqrt{1 + {\vec h}^2}\, (\tau + {{\vec h
 \cdot \vec z}\over {M}});  {{\vec z}\over {M}} + (\tau + {{\vec h
 \cdot \vec z}\over {M}})\, \vec h + {{{\vec {\bar S}} \times \vec
 h}\over {M\, (1 + \sqrt{1 + {\vec h}^2})}} \Big) = z_W^{\mu}(\tau
 ,\vec 0),\nonumber \\
 &&{}\nonumber \\
  x^{\mu}_o(\tau ) &{\buildrel {def}\over =}& x^{\mu}(\tau ,0) =
  Y^{\mu}(\tau ) + \epsilon^{\mu}_r(\vec h)\, \eta^r_o(\tau ),\qquad
  {\vec \eta}_o(\tau )\, {\buildrel {def}\over =}\, \vec \eta
  (\tau ,0),\nonumber \\
    x^{\mu}_{\pi}(\tau ) &{\buildrel {def}\over =}& x^{\mu}(\tau ,\pi ) =
  Y^{\mu}(\tau ) + \epsilon^{\mu}_r(\vec h)\, \eta^r_{\pi}(\tau ),\qquad
  {\vec \eta}_{\pi}(\tau )\, {\buildrel {def}\over =}\, \vec \eta
  (\tau ,\pi ),\nonumber \\
  &&{}\nonumber \\
  &&{\dot x}_o^2(\tau ) = 0, \rightarrow {\dot {\vec \eta}}^2_o =
  1,\qquad {\dot x}_{\pi}^2(\tau ) = 0, \rightarrow {\dot
  {\vec \eta}}^2_{\pi} =  1,
  \label{4.2}
  \eea

\noindent where the expression of the Fokker-Pryce center of
inertia, as a function of $\tau$, $\vec z$, $\vec h$, $M =
\sqrt{P^2}$ and ${\vec {\bar S}}$, was taken from Ref.\cite{4a}.

\medskip

This shows that inside the Wigner 3-spaces the open string is
described by the  Wigner spin-1 3-vectors ${\vec{\eta}}(\tau
,\lambda)$, ${\vec \kappa}(\tau ,\lambda)$ \footnote{Under Lorentz
transformations $\Lambda $ these 3-vectors rotate with Wigner
rotations [$\overset{\circ }{P}=M\,(1;\vec{0})$; $P^{\mu }=L(P,\overset{%
\circ }{P})^{\mu }{}_{\nu }\,\overset{\circ }{P}^{\nu }=\epsilon
_{A=\nu }^{\mu }(u(P))\,\overset{\circ }{P}^{\nu }$ with $L$ the
standard Wigner boost]\hfill \break
\begin{eqnarray*}
R^{\mu }{}_{\nu }(\Lambda ,P) &=&{[L(\overset{\circ }{P},P)\Lambda
^{-1}L(\Lambda P,\overset{\circ }{P})]}^{\mu }{}_{\nu }=\left(
\begin{array}{cc}
1 & 0 \\
0 & R^{i}{}_{j}(\Lambda ,P)
\end{array}
\right) , \\
{} &&{} \\
R^{i}{}_{j}(\Lambda ,P) &=&{(\Lambda
^{-1})}^{i}{}_{j}-{\frac{{(\Lambda ^{-1})^{i}{}_{o}\,P_{\beta
}\,(\Lambda ^{-1})^{\beta }{}_{j}}}{{P_{\rho
}\,(\Lambda ^{-1})^{\rho }{}_{o}+\sqrt{P^{2}}}}}- \\
&-&{\frac{{P^{i}}}{{P^{o}+\sqrt{P^{2}}}}}[(\Lambda
^{-1})^{o}{}_{j}-{\frac{{ ((\Lambda ^{-1})^{o}{}_{o}-1)\,P_{\beta
}\,(\Lambda ^{-1})^{\beta }{}_{j}}}{{ P_{\rho }\,(\Lambda
^{-1})^{\rho }{}_{o}+\sqrt{P^{2}}}}}].
\end{eqnarray*}
As a consequence the scalar product of two of these 3-vectors is a
Lorentz scalar.} as independent canonical variables. To them we must
add $P^{\mu }$ of Eq.(\ref{3.7}) and a canonically conjugate
collective variable ${\tilde{x}}^{\mu }$ and then replace them with
the final center-of-mass variables $\vec z$ and $\vec h$.
\medskip

It turns out  that the relevant collective variable to be added is
the \textit{external} canonical non-covariant 4-center of mass
${\tilde x}^{\mu}(\tau)$. From Ref.\cite{4a} we have the following
expressions for ${\tilde x}^{\mu}(\tau)$ in terms of $\tau$, $\vec
z$, $\vec h$, $M = \sqrt{P^2}$ and ${\vec {\bar S}}$, and for the
Lorentz generators $J^{\mu\nu}$ of Eq.(\ref{3.7}) (included the
definition of the rest-frame spin ${\vec {\bar S}}$)

\begin{eqnarray}
 {\tilde{x}}^{\mu }(\tau ) &=&
 \Big(\sqrt{1 + {\vec h}^2}\, (\tau +
 {{\vec h \cdot \vec z}\over {M}}); {{\vec z}\over {M}} + (\tau +
 {{\vec h \cdot \vec z}\over {M}})\, \vec h\Big) = \nonumber \\
 &=& z^{\mu}_W(\tau,{\tilde {\vec \sigma}}) = Y^{\mu}(\tau ) +
 \Big(0; {{ - {\vec {\bar S}} \times \vec h}\over {M\, (1 +
 \sqrt{1 + {\vec h}^2})}}\Big), \nonumber \\
 &&  \notag \\
 J^{\mu \nu } &=&[h^{\mu }\,\epsilon _{r}^{\nu }(\vec h) - h^{\nu}\,
 \epsilon _{r}^{\mu }(\vec h)]\, {\bar{S}}^{or} + \epsilon _{r}^{\mu
 }(\vec h)\, \epsilon _{s}^{\nu }(\vec h)\, {\bar{S}}^{rs},  \notag \\
 {\bar{S}}^{rs} &\equiv & \int^{\pi}_{o} d\lambda\,
 [\eta^{r}\, k^{s} - \eta^{s}\, k^{r}](\tau ,\lambda ),\nonumber \\
 {\bar{S}}^{or}&\equiv& -\, \int_{o}^{\pi} d\lambda\,
 \eta^r(\tau ,\lambda )\, \sqrt{ - N^2\, \Big(\partial_{\lambda}\, \vec \eta
 (\tau ,\lambda )\Big)^2  +
 {\vec \kappa}^2(\tau ,\lambda )},  \notag \\
 {\tilde{S}}^{ij} &=&\delta ^{ir}\,\delta ^{js}\,{\bar{S}}^{rs},\quad
 \quad { \tilde{S}}^{oi}=-{\frac{{\delta
 ^{ir}\,{\bar{S}}^{rs}\,P^{s}}}{{P^{o}+M}}}.
 \label{4.3}
\end{eqnarray}

\medskip

The external Poincare' generators are [while $i,j..$ are Euclidean
indices, $r,s..$ are Wigner spin-1 indices; ${\tilde S}^{\mu\nu}$ is
given in Eq.(\ref{4.3})]

\begin{eqnarray}
 P^{\mu } &,&\qquad J^{\mu \nu }=x_{s}^{\mu }\,P^{\nu }-x_{s}^{\nu
 }\,P^{\mu }+S^{\mu \nu }={\tilde{x}}^{\mu }\,P^{\nu
 }-{\tilde{x}}^{\nu }\,P^{\mu }+{\tilde{S}}^{\mu \nu },  \notag \\
 &&{}  \notag \\
 P^{o} &=&\sqrt{M^{2}+{\vec{P}}^{2}} = M\, \sqrt{1 + {\vec h}^2},\quad
 \vec P = M\, \vec h,  \notag \\
 &&{}  \notag \\
 J^{ij} &=& {\tilde{x}}^{i}\,P^{j}-{\tilde{x}}^{j}\,P^{i}+\delta
 ^{ir}\,\delta^{js}\,\epsilon ^{rsu}\,{\bar{S}}^{u} = z^i\, h^j -
 z^j\, h^i + \epsilon^{iju}\, {\bar S}^u,  \notag \\
 K^{i}&=&J^{oi}= {\tilde{x}}^{o}\,P^{i}-{\tilde{x}}^{i}\,
 \sqrt{M^{2}+{\vec{P}}^{2}}-{\frac{{\delta ^{ir}\,P^{s}\,\epsilon
 ^{rsu}\,{\bar{S}}^{u}}}{{M+\sqrt{M^{2}+{\vec{P }}^{2}}}}} =\nonumber \\
 &=& - \sqrt{1 + {\vec h}^2}\, z^i + {{({\vec {\bar S}} \times \vec h)^i}\over
 {1 + \sqrt{1 + {\vec h}^2}}}.
  \label{4.4}
\end{eqnarray}

Note that both ${\tilde L}^{\mu\nu}={\tilde x}^{\mu}\, P^{\nu} -
{\tilde x} ^{\nu}\, P^{\mu}$ and ${\tilde S}^{\mu\nu}$ are
conserved.

\bigskip

Inside the Wigner 3-spaces the Poincare' generators (\ref{3.7}) take
the following form

 \bea
 P^{\mu} &\approx& h^{\mu}\, \int_o^{\pi} d\lambda\,
 \sqrt{-N^2\, {\vec \eta}{}^{\,\prime\, 2}(\tau ,\lambda ) +
 {\vec \kappa}^2(\tau ,\lambda )} + N\,
 \epsilon^{\mu}_r(\vec h)\, \int_o^{\pi} d\lambda\, \kappa^r(\tau
 ,\lambda ) \approx\nonumber \\
  &\approx& h^{\mu}\, \int_o^{\pi} d\lambda\,
  \sqrt{-N^2\, {\vec \eta}{}^{\,\prime\, 2}(\tau ,\lambda ) +
 {\vec \kappa}^2(\tau ,\lambda )},\nonumber \\
 J^{\mu\nu} &\approx& \Big[h^{\mu}\, \epsilon^{\nu}_r(\vec h) - h^{\nu}\,
 \epsilon^{\mu}_r(\vec h)\Big]\, \Big[\tau\, \int_o^{\pi} d\lambda\, \kappa^r(\tau
 ,\lambda  ) -\nonumber \\
 &-& \int_o^{\pi} d\lambda\, \eta^r(\tau ,\lambda )\,
 \sqrt{-N^2\, {\vec \eta}{}^{\,\prime\, 2}(\tau ,\lambda ) +
 {\vec \kappa}^2(\tau ,\lambda )}  \Big] +\nonumber \\
 &+& \Big[\epsilon^{\mu}_r(\vec h)\, \epsilon^{\nu}_s(\vec h) - \epsilon^{\nu}_r(\vec h)\,
 \epsilon^{\mu}_s(\vec h)\Big]\, \int_o^{\pi} d\lambda\, \eta^r(\tau ,\lambda )\,
 \kappa^s(\tau ,\lambda ) \approx \nonumber \\
 &\approx& \Big[h^{\mu}\, \epsilon^{\nu}_r(\vec h) - h^{\nu}\,
 \epsilon^{\mu}_r(\vec h)\Big]\, \Big[- \int_o^{\pi} d\lambda\, \eta^r(\tau ,\lambda )\,
 \sqrt{-N^2\, {\vec \eta}{}^{\,\prime\, 2}(\tau ,\lambda ) +
 {\vec \kappa}^2(\tau ,\lambda )}  \Big] +\nonumber \\
 &+& \Big[\epsilon^{\mu}_r(\vec h)\, \epsilon^{\nu}_s(\vec h) - \epsilon^{\nu}_r(\vec h)\,
 \epsilon^{\mu}_s(\vec h)\Big]\, \int_o^{\pi} d\lambda\, \eta^r(\tau ,\lambda )\,
 \kappa^s(\tau ,\lambda ),
 \label{4.5}
 \eea

As a consequence, inside the Wigner hyper-plane there is the
following unfaithful realization of the Poincare' algebra, whose
inner generators are

\bea
 M &=& \int_0^{\pi} d\lambda\, \sqrt{N^2\, \Big(\partial_{\lambda}\, \vec \eta
 (\tau ,\lambda )\Big)^2 + {\vec \kappa}^2(\tau ,\lambda )},\nonumber \\
 \vec k &=& \int^{\pi}_{-\pi} d\lambda\, \vec \kappa(\tau
 ,\lambda ) \approx 0,\nonumber \\
 {\vec {\cal J}} &=& {\vec {\bar S}} = \int^{\pi}_{o}
 d\lambda\, \vec \eta (\tau ,\lambda ) \times \vec \kappa (\tau
 ,\lambda ),\nonumber \\
 {\cal K}^r &=& {\bar S}^{or} = -\int_{o}^{\pi} d\lambda\,
 \eta^r(\tau ,\lambda )\, \sqrt{N^2\, \Big(\partial_{\lambda}\, \vec \eta
 (\tau ,\lambda )\Big)^2  +
 {\vec \kappa}^2(\tau ,\lambda )} \approx 0.
 \label{4.6}
 \eea
\medskip

As already said, while $\vec k \approx 0$ is the rest-frame
condition, ${\cal K}^r \approx 0$ eliminates the inner center of
mass inside the Wigner hyper-planes and, as shown in Ref.\cite{4a},
implies that the embedding (\ref{4.1}), defining them, is centered
on the inertial covariant non-canonical Fokker-Pryce center of
inertia $Y^{\mu}(\tau)$. The rest-frame condition $\vec k \approx 0$
implies ${\dot{x}}_{s}^{\mu }(\tau ) = {\dot{\tilde{x}}} ^{\mu
}(\tau ) = {\dot Y}^{\mu}(\tau) = u^{\mu }(P)$, i.e. the velocities
are all parallel to $ P^{\mu }$, so that there is no
\textit{classical zitterbewegung}.

\bigskip

We must now find a  canonical basis of relative variables inside the
Wigner 3-spaces by defining a rest-frame instant form analogue of
Eqs. (\ref{2.25})-(\ref{2.27}). This can be done with the following
canonical transformation

\begin{eqnarray*}
 &&\begin{minipage}[t]{3cm}
\begin{tabular}{|l|} \hline
$\vec \eta (\tau ,\lambda )$ \\ \hline $\vec k(\tau ,\lambda )$
\\ \hline
\end{tabular}
\end{minipage} \hspace{1cm} {\longrightarrow \hspace{.2cm}} \
\begin{minipage}[t]{4 cm}
\begin{tabular}{|l|l|} \hline
$\vec \eta (\tau )$ & $\vec y(\tau ,\lambda )$ \\
\hline  $\vec k(\tau ) \approx 0$ & ${\vec {\cal P}}(\tau ,\lambda
)$\\ \hline
\end{tabular}
\end{minipage},
 \end{eqnarray*}

\begin{eqnarray*}
 \vec \eta (\tau ) &=& {1\over {2\pi}}\, \int^{\pi}_{-\pi}
 d\lambda\, \vec \eta (\tau ,\lambda ),\qquad \vec k = {1\over
 2}\, \int^{\pi}_{-\pi} d\lambda\, \vec \kappa (\tau ,\lambda ) \approx
 0,\nonumber \\
 \vec y (\tau ,\lambda ) &=& - \partial_{\lambda}\, \vec \eta
 (\tau ,\lambda ),\qquad {\vec {\cal P}}(\tau ,\lambda ) =
 \int_o^{\lambda} d\lambda_1\, \vec \kappa(\tau ,\lambda_1) -
 {{\lambda}\over {\pi}}\, \vec k,\nonumber \\
 &&\int^{\pi}_{-\pi} d\lambda\, \vec y(\tau ,\lambda ) =
 \int^{\pi}_{-\pi} d\lambda\, {\vec {\cal P}}(\tau ,\lambda ) =
 0,\nonumber \\
 &&{\vec {\cal P}}(\tau ,0) = {\vec {\cal P}}(\tau ,\pm \pi ) =0,
 \Rightarrow \int^{\pi}_{-\pi} d\lambda\, \partial_{\lambda}\,
 {\vec {\cal P}}(\tau ,\lambda ) = 0,\nonumber \\
 &&{}\nonumber \\
 &&\{ \eta^i, k^j \} = \delta^{ij},\qquad \{ y^i(\tau ,\lambda ),
 {\cal P}^j(\tau ,\lambda_1) \} = \delta^{ij}\,
 \triangle_{-}(\lambda ,\lambda_1),
 \end{eqnarray*}

\bea
 \vec \eta (\tau ,\lambda ) &=& \vec \eta (\tau ) + {1\over
 {2\pi}}\, \int^{\pi}_{-\pi} d\lambda_1\, \int_\lambda^{\lambda_1}
 d\lambda_2\, \vec y(\tau ,\lambda_2) \nonumber \\
 &=&  \vec\eta (\tau ) + \vec \zeta (\tau ,\lambda ),\nonumber \\
 \vec \kappa(\tau ,\lambda ) &=& {{\vec k}\over {\pi}} +
 \partial_{\lambda}\, {\vec {\cal P}}(\tau ,\lambda ) \approx
 \partial_{\lambda}\, {\vec {\cal P}}(\tau ,\lambda ),\nonumber \\
 &&{}\nonumber \\
  &&{\vec \eta}_o(\tau ) = \vec \eta (\tau ,0) = \vec \eta (\tau )
 + {1\over {2\pi}}\, \int^{\pi}_{-\pi} d\lambda_1\,
 \int_o^{\lambda_1} d\lambda_2\, \vec y(\tau ,\lambda_2),\nonumber \\
 &&{\vec \eta}_{\pi}(\tau ) = \vec \eta (\tau , \pi) = {\vec \eta}_o(\tau )
 -  \int^{\pi}_o d\lambda_2\, \vec y(\tau ,\lambda_2),\qquad
 {\dot {\vec \eta}}_o^2(\tau ) = {\dot {\vec \eta}}_\pi^2(\tau )
 = 1,\nonumber \\
 &&{}\nonumber \\
 &&\vec S \approx \int^{\pi}_{-\pi} d\lambda\, \vec
 \zeta (\tau ,\lambda ) \times \partial_{\lambda}\, {\vec {\cal
 P}}(\tau ,\lambda ) = \int^{\pi}_{o} d\lambda\,
 \vec y(\tau ,\lambda ) \times {\vec {\cal P}}(\tau ,\lambda ).
 \label{4.7}
 \eea

\bigskip

In the new canonical basis the inner Poincare' generators and the
first class constraints take the following form [in accord with
Eq.(\ref{3.6})]

\bea
 M &\approx& \int_o^{\pi} d\lambda\,
 \sqrt{-N^2\, {\vec y}^2(\tau ,\lambda)   + \Big( {\partial_\lambda \vec {\cal P}}(\tau
 ,\lambda )\Big)^2},\nonumber \\
 &&\nonumber\\
 {\cal K}^i &\approx&
 - \eta^i(\tau )\, \int^{\pi}_o d\lambda\,
 \sqrt{-N^2\, {\vec y}^2(\tau ,\lambda)   + \Big( {\partial_\lambda \vec {\cal P}}(\tau
 ,\lambda )\Big)^2} -\nonumber \\
 &-& {1\over {2\pi}}\,\int^{\pi}_o d\lambda\,
 \int^{\pi}_{-\pi} d\lambda_1\,
 \int_\lambda^{\lambda_1} d\lambda_2\, y^i(\tau ,\lambda_2)\,
 \sqrt{-N^2\, {\vec y}^2(\tau ,\lambda)   + \Big( {\partial_\lambda \vec {\cal P}}(\tau
 ,\lambda )\Big)^2} \approx 0,\nonumber \\
 &&\nonumber\\
  \vec k &\approx& 0,\nonumber \\
  &&\nonumber\\
 \vec S &\approx &  \int_{o}^{\pi} d\lambda\,
 \vec y(\tau ,\lambda ) \times {\vec {\cal P}}(\tau ,\lambda ),\nonumber \\
 &&{}\nonumber \\
 \chi (\tau ,\lambda ) &=& \vec \kappa(\tau ,\lambda ) \cdot
 \partial_{\lambda}\, \vec \eta (\tau ,\lambda ) \approx
 - \partial_{\lambda}\, {\vec {\cal P}}(\tau ,\lambda ) \cdot
 \vec y(\tau ,\lambda ) \approx 0,\nonumber \\
 &&{}\nonumber \\
 \{ \chi (\tau ,\lambda ), \chi (\tau ,\lambda_1) \} &=& +\chi
 (\tau ,\lambda )\, \partial_{\lambda}\, \delta (\lambda -
 \lambda_1) - \chi (\tau ,\lambda_1)\, \partial_{\lambda_1}\,
 \delta (\lambda - \lambda_1) =\nonumber \\
 &=& + \chi (\tau ,\lambda )\, {{\partial \Delta_-(\lambda ,\lambda_1)}
 \over {\partial \lambda}} -
 \chi (\tau ,\lambda_1)\, {{\partial \Delta_{+}(\lambda ,\lambda_1)}\over
 {\partial \lambda_1}}.
 \label{4.8}
 \eea

\medskip

The conditions ${\cal K}^r \approx 0$ eliminating the inner center
of mass give the following determination of the collective variable
$\vec \eta(\tau)$

\bea
 \vec \eta (\tau ) &\approx& +
{1\over { 2\pi\,\int_o^{\pi} d\lambda \, \sqrt{-N^2\, {\vec
y}^2(\tau ,\lambda)   + \Big( {\partial_\lambda \vec {\cal P}}(\tau
,\lambda )\Big)^2} }} \nonumber \\
 &&  \int_o^{\pi} d\lambda\,
 \int^{\pi}_{-\pi} d\lambda_1\, \int_\lambda^{\lambda_1}\,
 \vec y(\tau ,\lambda_2)\,
 \sqrt{-N^2\, {\vec y}^2(\tau ,\lambda)   + \Big( {\partial_\lambda \vec
 {\cal P}}(\tau ,\lambda )\Big)^2},\nonumber \\
 &&{}\nonumber \\
 &&\Downarrow\nonumber \\
 &&{}\nonumber \\
 \vec \eta (\tau ,\lambda ) &\approx& {1\over {2\pi}}\,
 \int^{\pi}_{-\pi} d\lambda_1\, \int_\lambda^{\lambda_1} d\lambda_2\,
 \vec y(\tau ,\lambda_2) +
{1\over { 2\pi\,\int_o^{\pi} d\lambda \, \sqrt{-N^2\, {\vec
y}^2(\tau ,\lambda)   + \Big( {\partial_\lambda \vec {\cal P}}(\tau
,\lambda )\Big)^2}
 }} \nonumber \\
 &&  \int_o^{\pi} d\lambda\,
 \int^{\pi}_{-\pi} d\lambda_1\, \int_\lambda^{\lambda_1}\,
 \vec y(\tau ,\lambda_2)\,
 \sqrt{-N^2\, {\vec y}^2(\tau ,\lambda)   + \Big( {\partial_\lambda
 \vec {\cal P}}(\tau ,\lambda )\Big)^2},\nonumber \\
 &&{}\nonumber \\
 x^{\mu}(\tau, \lambda) &=& Y^{\mu}(\tau) + \epsilon^{\mu}_r(\vec h)\,
 \eta^r(\tau, \lambda).\nonumber \\
 &&{}
 \label{4.9}
 \eea

\bigskip

Since we have $\{ \vec k, \chi (\tau ,\lambda ) \} = 0$, the gauge
fixings ${\vec {\cal K}} \approx 0$ lead to the Dirac bracket $\{
\chi (\tau ,\lambda ), \chi (\tau ,\lambda^{'}) \}^* = \{ \chi (\tau
,\lambda ), \chi (\tau ,\lambda^{'}) \}$, namely Eqs.(\ref{4.8})
remain unchanged.

\vfill\eject

\section{The Frenet-Serret canonical basis}

The next step is to find a canonical transformation from the
canonical basis (\ref{4.7}) for the relative variables to a new
basis also adapted to the transversality constraints given in
Eqs.(\ref{4.8}). This can be done by using a Frenet-Serret (FS)
description of the vectors tangent and normal to the string (see
Ref.\cite{23a}).
\medskip

The unit tangent $\hat t(\tau ,\lambda )$, ${\hat t}^2(\tau ,\lambda
) = 1$, to the string inside the Wigner hyper-plane and the final
expression of the transversality constraint and of the invariant
mass are

\bea
 \hat t(\tau ,\lambda) &=& - {{\vec y(\tau,\lambda)}\over {h(\tau ,\lambda)}}
  = {{\partial_{\lambda}\, \vec \eta (\tau ,\lambda)}\over {h(\tau ,\lambda)}},
  \nonumber \\
 &&{}\nonumber \\
 \Rightarrow && h(\tau ,\lambda) = \sqrt{{\vec y}^2(\tau ,\lambda)}
 {\buildrel {def}\over =} \partial_{\lambda}\, s(\tau
 ,\lambda),\qquad \{ s(\tau, \lambda), s(\tau, \lambda_1) \} = 0,\nonumber \\
 &&{}\nonumber \\
 \Rightarrow && \chi (\tau ,\lambda ) = h(\tau ,\lambda )\,
 \partial_{\lambda}\, {\vec {\cal P}}(\tau ,\lambda ) \cdot \hat
 t(\tau ,\lambda ) \approx 0,\nonumber \\
 &&{}\nonumber \\
 or && \pi_{\chi}(\tau ,\lambda ) {\buildrel {def}\over =}\, + {{\chi (\tau
 ,\lambda )}\over {h(\tau ,\lambda )}} =  \Big[\partial_{\lambda}\,
 {\vec {\cal P}} \cdot \hat t\Big](\tau ,\lambda) \approx 0,\nonumber \\
 &&{}\nonumber \\
 M &=&  \int_o^{\pi} d\lambda\, \sqrt{-N^2\,
 (\partial_{\lambda}\, s(\tau ,\lambda ))^2 + \Big(\partial_\lambda\vec {\cal P}(\tau ,\vec \lambda
 )\Big)^2}.
 \label{5.1}
 \eea

The density of invariant mass depends upon the gauge variable
$s(\tau ,\lambda )$, which describes the arc-length along the string
\footnote{If we invert $s = s(\tau, \lambda)$ to $\lambda =
\lambda(\tau, s)$, we get $\hat t(\tau, \lambda(\tau, s)) =
{{\partial_{\lambda}\, \vec \eta(\tau, \lambda)}\over
{\partial_{\lambda}\, s(\tau, \lambda)}} = {{\partial\, \vec
\eta(\tau, \lambda(\tau, s))}\over {\partial\, s}}$.}.\medskip

In the gauge $s(\tau ,\lambda ) = \lambda$, $h(\tau ,\lambda ) = 1$,
$\vec y(\tau ,\lambda ) = -\hat t(\tau ,\lambda )$ we have
\[M =\int_o^{\pi} d\lambda\,
\sqrt{-N^2 + \Big(\partial_\lambda\vec {\cal P}(\tau ,\vec \lambda
)\Big)^2}:\] let us call this gauge the {\it FS gauge}.
\medskip

\bigskip
Since we have

\bea
 \{ y^r(\tau ,\lambda ), \chi (\tau ,\lambda_1) \} &=&
 \partial_{\lambda_1}\, \eta^r(\tau ,\lambda_1 )\,
 \partial_{\lambda}\, \delta (\lambda - \lambda_1) = +
 \partial_{\lambda}\, \Big[y^r(\tau ,\lambda )\, \delta (\lambda
 - \lambda_1)\Big],\nonumber \\
 &=&\partial_{\lambda_1}\, \eta^r(\tau ,\lambda_1 )\,
 \partial_{\lambda}\, \Delta_+ (\lambda,\lambda_1) = +
 \partial_{\lambda}\, \Big[y^r(\tau ,\lambda )\, \Delta_-
 (\lambda,\lambda_1)\Big],\nonumber \\
 &&{}\nonumber \\
 \{ h(\tau ,\lambda ), \chi (\tau ,\lambda_1) \} &=& +
 \partial_{\lambda}\, \Big[h(\tau ,\lambda )\, \delta (\lambda
 - \lambda_1)\Big]=+\partial_{\lambda}\, \Big[h(\tau ,\lambda )\,
 \Delta_- (\lambda,\lambda_1)\Big],\nonumber \\
 &&{}\nonumber \\
  \{ s(\tau ,\lambda ), \chi (\tau ,\lambda_1) \} &=&
 + h(\tau ,\lambda )\, \delta (\lambda - \lambda_1) = +
 \partial_{\lambda}\, s(\tau ,\lambda )\, \delta (\lambda -
 \lambda_1)=\nonumber\\
 &=&
 + h(\tau ,\lambda )\, \Delta_- (\lambda , \lambda_1) = +
 \partial_{\lambda}\, s(\tau ,\lambda )\, \Delta_- (\lambda ,
 \lambda_1),
 \label{5.2}
 \eea

 \noindent we get an Abelianization of the transversality
 constraints with the new Poisson brackets

\bea
  \{ s(\tau ,\lambda ), \pi_{\chi}(\tau ,\lambda_1) \} &=&
 \delta (\lambda - \lambda_1) = \Delta_{-}(\lambda ,\lambda_1),\nonumber \\
 &&{}\nonumber \\
 \{ \pi_{\chi}(\tau ,\lambda ), \pi_{\chi}(\tau ,\lambda_1) \}
 &=& 0,\qquad \{ s(\tau, \lambda), s(\tau, \lambda_1) \} = 0.
 \label{5.3}
 \eea

\bigskip

On the Euclidean Wigner 3-space $\Sigma_{\tau}$ the open string is
described by the Wigner 3-vector $\vec \eta(\tau, \lambda)$ with
$\lambda \in (0, \pi)$. Let $\hat t(\tau ,\lambda )$, $\hat n(\tau
,\lambda )$ and $\hat b(\tau ,\lambda )$ be the unit tangent, unit
normal and unit binormal 3-vectors to the string $\vec \eta (\tau
,\lambda )$ inside the Wigner hyper-plane. Since $s = s(\tau,
\lambda)$ is the arc-length along the string, the FS formulas are
\begin{eqnarray*}
{{\partial_{\lambda}\, \hat t(\tau, \lambda)}\over
{\partial_{\lambda}\, s(\tau, \lambda)}} &=& \rho_1(\tau, \lambda)\,
\hat n(\tau, \lambda)\nonumber\\
&&\nonumber\\
{{\partial_{\lambda}\, \hat n(\tau, \lambda)}\over
{\partial_{\lambda}\, s(\tau, \lambda)}} &=& - \rho_1(\tau,
\lambda)\, \hat t(\tau, \lambda) + \rho_2(\tau,
\lambda)\, \hat b(\tau, \lambda)\nonumber\\
&&\nonumber\\
{{\partial_{\lambda}\, \hat b(\tau, \lambda)}\over
{\partial_{\lambda}\, s(\tau, \lambda)}} &=& - \rho_2(\tau,
\lambda)\, \hat n(\tau, \lambda),
\end{eqnarray*}
where $\rho_1(\tau, \lambda)$ is the {\it curvature} of the string
and $\rho_2(\tau, \lambda)$ its {\it torsion}. These two quantities
are Wigner 3-scalars and therefore {\it Lorentz scalars}.

\medskip

The tangent vector, defined in Eq.(\ref{5.1}), can be parametrized
with two angles $\theta(\tau, \lambda)$, $\phi(\tau, \lambda)$,
allowing one to find the following ortho-normal triad $\Big(\hat
t(\tau, \lambda), {\hat b}_{\theta}(\tau, \lambda), {\hat
b}_{\phi}(\tau, \lambda)\Big)$

\bea
 \hat t(\tau ,\lambda ) &=& - {{\vec y(\tau ,\lambda )}\over
 {h(\tau ,\lambda )}} {\buildrel {def}\over =}
 \Big(\sin \theta \, \cos \phi (\tau ,\lambda
 ), \sin \theta \, \sin \phi ,
 \cos \theta  \Big)(\tau ,\lambda ),\nonumber \\
 &&{}\nonumber \\
   {\hat b}_{\theta}(\tau ,\lambda ) &=& {{\partial\, \hat t(\tau
  ,\lambda )}\over {\partial\, \theta}} = \Big(\cos \theta\, \cos
  \phi , \cos \theta\, \sin \phi , - \sin \theta\Big)(\tau ,\lambda
  ),\nonumber \\
  &&\nonumber\\
 {\hat b}_{\phi}(\tau ,\lambda ) &=& \Big({1\over {\sin \theta}}\,
 {{\partial\, \hat t}\over {\partial\, \phi}}\Big)(\tau ,\lambda ) =
 \Big(- \sin \phi , \cos \phi ,0\Big)(\tau ,\lambda ),
 \label{5.4}
 \eea

\medskip

The first FS equation becomes \footnote{We also have $
\partial_{\lambda}\, {\hat b}_{\theta}(\tau ,\lambda ) =
\Big[- \partial_{\lambda}\,\theta\,  \hat t + \cos \theta\,
\partial_{\lambda}\, \phi\, {\hat b}_{\phi}\Big]
(\tau ,\lambda )$ and $\partial_{\lambda}\, {\hat b}_{\phi}(\tau
,\lambda ) = - \Big[\partial_{\lambda}\, \phi\, \Big(\sin \theta\,
 \hat t + \cos \theta\, {\hat b}_{\theta}\Big)\Big](\tau ,\lambda
 )$.}

\begin{eqnarray*}
 && \partial_{\lambda}\, \hat t(\tau ,\lambda ) = \partial_{\lambda}\,
 s(\tau ,\lambda )\, \rho_1(\tau ,\lambda )\, \hat n(\tau ,\lambda )
 =\nonumber \\
 &&\nonumber\\
 &=&\Big( \cos \theta\, \cos \phi\, \partial_{\lambda} \theta -
 \sin \theta\, \sin \phi\, \partial_{\lambda}\, \phi ,\cos \theta\, \sin \phi\, \partial_{\lambda}\, \theta +
 \sin \theta\, \cos \phi\, \partial_{\lambda}\, \phi ,
  - \sin \theta\, \partial_{\lambda}\, \theta \Big)(\tau ,\lambda
  ) =  \nonumber \\
  &&\nonumber\\
  &=& \Big[{{\partial \hat t}\over {\partial
 \theta}}\, \partial_{\lambda}\, \theta + {{\partial \hat t}\over
 {\partial \phi}}\, \partial_{\lambda}\, \phi\Big](\tau ,\lambda ) =
 \Big[\partial_{\lambda}\, \theta\, {\hat b}_{\theta} +
 \sin \theta\, \partial_{\lambda}\, \phi\, {\hat b}_{\phi}\Big]
 (\tau ,\lambda ),\nonumber \\
 \end{eqnarray*}

\begin{eqnarray}
&&\Downarrow\nonumber \\
 &&{}\nonumber \\
\rho_1(\tau ,\lambda ) &=& {1\over {\partial_{\lambda}\, s(\tau
,\lambda )}}\,\sqrt{\Big(\partial_{\lambda}\, \hat t(\tau ,\lambda
)\Big)^2}
 = {1\over {\partial_{\lambda}\, s(\tau ,\lambda )}}\, \sqrt{(\partial_{\lambda}\, \theta)^2 +
 \sin^2 \theta\, (\partial_{\lambda}\, \phi)^2}(\tau ,\lambda
 )\nonumber\\
 &&\nonumber\\
 &&{}\nonumber \\
\hat n(\tau, \lambda) &=& {{[\partial_{\lambda}\, \theta\, {\hat
b}_{\theta} +
 \sin \theta\, \partial_{\lambda}\, \phi\, {\hat b}_{\phi}](\tau,
 \lambda)}\over {\partial_{\lambda}\, s(\tau, \lambda)\,\,
 \rho_1(\tau, \lambda)}} ={{[\partial_{\lambda}\, \theta\, {\hat b}_{\theta} +
 \sin \theta\, \partial_{\lambda}\, \phi\, {\hat b}_{\phi}]}\over {\sqrt{(\partial_{\lambda}\, \theta)^2 +
 \sin^2 \theta\, (\partial_{\lambda}\, \phi)^2}}}\,(\tau,
 \lambda)
 \label{5.5}
\end{eqnarray}

Moreover we have

\begin{eqnarray}
\hat{b}(\tau,
\lambda)&=&\hat{t}(\tau,\lambda)\times\hat{n}(\tau,\lambda)=
 {{[\partial_{\lambda}\, \theta\, {\hat b}_{\phi} -
 \sin \theta\, \partial_{\lambda}\, \phi\, {\hat b}_{\theta}]}\over
 {\sqrt{(\partial_{\lambda}\, \theta)^2 +
 \sin^2 \theta\, (\partial_{\lambda}\, \phi)^2}}}\,(\tau,
 \lambda)\nonumber\\
 &&\nonumber\\
&&\Downarrow\nonumber \\
 &&{}\nonumber \\
 \partial_\lambda\hat{b}(\tau,\lambda)&=&
 {{\sin \theta\, (\partial_{\lambda}\, \phi\, \partial^2_{\lambda}\, \theta
 - \partial_{\lambda}\, \theta\, \partial^2_{\lambda}\, \phi) -
 \sin^2 \theta\, \cos \theta\, (\partial_{\lambda}\, \phi)^3 - 2\,
 \cos \theta\, (\partial_{\lambda}\, \theta )^2\, \partial_{\lambda}\, \phi }\over
 {\Big[(\partial_{\lambda}\, \theta)^2 + \sin^2 \theta\,
 (\partial_{\lambda}\, \phi)^2\Big]^{3/2}}}(\tau,
 \lambda)\times\nonumber\\
 &&\nonumber\\
 &&\qquad\times[\partial_{\lambda}\, \theta\, {\hat b}_{\theta} +
 \sin \theta\, \partial_{\lambda}\, \phi\, {\hat b}_{\phi}](\tau,
 \lambda)\nonumber\\
 &&\nonumber\\
&&\Downarrow\nonumber \\
 &&{}\nonumber \\
 \rho_2(\tau ,\lambda ) &=&{1\over {\partial_{\lambda}\, s(\tau ,\lambda )}}\,
 \sqrt{\Big(\partial_{\lambda}\, \hat b(\tau ,\lambda )\Big)^2} =\nonumber \\
 &&\nonumber\\
 &=& {{\sin \theta\, (\partial_{\lambda}\, \phi\, \partial2_{\lambda}\, \theta
 - \partial_{\lambda}\, \theta\, \partial^2_{\lambda}\, \phi) -
 \sin^2\, \theta\, \cos \theta\, (\partial_{\lambda}\, \phi)^3 - 2\,
 \cos \theta\, (\partial_{\lambda}\, \theta )^2\, \partial_{\lambda}\, \phi }\over
 {\partial_{\lambda}\, s \Big[(\partial_{\lambda}\, \theta)^2 + \sin^2\, \theta\,
 (\partial_{\lambda}\, \phi)^2\Big]}}(\tau ,\lambda ).\nonumber \\
 &&{}
  \label{5.6}
 \end{eqnarray}

\medskip

Let us remark that the tangents at the end points are determined by
$\vec y(\tau ,0/\pi ) = - \partial_{\lambda}\, s(\tau ,\lambda
){|}_{\lambda = o/\pi}\, \hat t(\tau ,0/\pi )$. Therefore, we have
$\int_o^{\pi} d\lambda\, \partial_{\lambda}\, \vec y(\tau ,\lambda )
= \vec y(\tau ,\pi ) - \vec y(\tau ,0)$, $\int_o^{\tau} d\lambda\,
\partial^2_{\lambda}\, \vec y(\tau \lambda ) =
\partial_{\lambda}\, \vec y(\tau ,\lambda ){|}_{\lambda = \pi} -
\partial_{\lambda}\, \vec y(\tau ,\lambda ){|}_{\lambda = 0}$.

\bigskip

Let us also remark that the negative parity of the variable $\vec
y(\tau, \lambda) = - \vec y(\tau, - \lambda)$ implies $\hat t(\tau,
- \lambda) = - \hat t(\tau, \lambda)$, so that from Eqs.(\ref{5.4})
we get $\cos\theta(\tau,-\lambda) = - cos\theta(\tau,\lambda)$. But
this implies $\theta(\tau, - \lambda) = \theta(\tau, \lambda) +
\pi$. Due to these discontinuities the description using the angles
$\theta(\tau, \lambda)$ and $\phi(\tau, \lambda)$ must be restricted
to the interval $(0, \pi)$, so that the distributions
$\triangle_{\pm}(\lambda, \lambda^{'})$ will be restricted to
$\delta(\lambda - \lambda^{'})$ from now on.

\bigskip

The previous geometrical description of the string suggests to
replace the canonical basis of relative variables $\vec y(\tau,
\lambda)$, ${\vec {\cal P}}(\tau, \lambda)$, with a new canonical
basis defined by the following point canonical transformation

\bea
 &&\begin{minipage}[t]{3cm}
\begin{tabular}{|l|} \hline
$\vec y (\tau ,\lambda )$ \\ \hline ${\vec {\cal P}}(\tau ,\lambda
)$\\ \hline
\end{tabular}
\end{minipage} \hspace{1cm} {\longrightarrow \hspace{.2cm}} \
\begin{minipage}[t]{4 cm}
\begin{tabular}{|l|l|l|} \hline
$s(\tau ,\lambda)$ & $\theta (\tau ,\lambda )$ & $\phi (\tau ,\lambda )$\\
\hline  $\pi_s(\tau ,\lambda )$ &
$\pi_{\theta}(\tau ,\lambda )$ & $\pi_{\phi}(\tau ,\lambda )$\\
\hline
\end{tabular}
\end{minipage},\nonumber \\
 &&{}\nonumber \\
 &&{}\nonumber \\
 &&\{ s(\tau ,\lambda ), \pi_s(\tau ,\lambda_1)\} = \{ \theta
 (\tau ,\lambda ), \pi_{\theta}(\tau ,\lambda_1)\} = \{ \phi (\tau
 ,\lambda ), \pi_{\phi}(\tau ,\lambda_1)\} = \delta(\lambda-\lambda_1).
 \label{5.7}
 \eea

\bigskip

The  point canonical transformation (\ref{5.7}) (the old momenta are
linearly connected to the new one) has the following generating
function

\bea
 \Phi &=& \int d\lambda\, {\cal P}^r(\tau ,\lambda )\, \Big(-
 \partial_{\lambda}\, s(\tau ,\lambda )\, {\hat t}^r(\tau ,\lambda )\Big)
 =\nonumber \\
 &&\nonumber\\
 &=& - \int d\lambda\, \partial_{\lambda}\, s(\tau ,\lambda )\,
 \Big[{\cal P}^1\, \sin \theta\, \cos \phi + {\cal P}^2\, \sin \theta\,
 \sin \phi + {\cal P}^3\, \cos \theta\Big](\tau ,\lambda ).
 \label{5.8}
 \eea

 The equations

 \bea
  \pi_s(\tau ,\lambda ) &=& {{\delta \Phi}\over {\delta s(\tau
  ,\lambda )}},\nonumber \\
    &&\nonumber\\
 \pi_{\theta}(\tau ,\lambda ) &=& {{\delta \Phi}\over {\delta \theta (\tau
  ,\lambda )}}, \nonumber \\
  &&\nonumber\\
    \pi_{\phi}(\tau ,\lambda ) &=& {{\delta \Phi}\over {\delta \phi (\tau
  ,\lambda )}},
 \label{5.9}
 \eea

\noindent imply

\bea
 \pi_s &=&  \partial_{\lambda}\, \Big({\vec {\cal P}} \cdot \hat
 t\Big),\nonumber \\
 &&\nonumber\\
 \pi_{\theta} &=& - {\cal P}^1\, (\partial_{\lambda}\, s)\, \cos
 \theta\, \cos \phi - {\cal P}^2\, (\partial_{\lambda}\, s)\,
 \cos \theta\, \sin \phi + {\cal P}^3\, \sin \theta = - \partial_{\lambda}\, s\, ({\vec {\cal P}} \cdot {\hat
 b}_{\theta}),\nonumber \\
 &&\nonumber\\
 \pi_{\phi} &=& {\cal P}^1\, (\partial_{\lambda}\, s)\, \sin
 \theta\, \sin \phi - {\cal P}^2\, (\partial_{\lambda}\, s)\,
 \sin \theta\, \cos \phi  = - \partial_{\lambda}\, s\, \sin \theta\, ({\vec {\cal P}} \cdot {\hat
 b}_{\phi}),
 \label{5.10}
 \eea

\noindent so that the expression of the old momenta in terms of the
new ones is

 \beq
 {\vec {\cal P}}(\tau ,\lambda ) =  \hat t(\tau ,\lambda )\,\int_o^{\lambda} d\lambda_1\,
 \pi_s(\tau ,\lambda_1)\,  - {1\over
 {\partial_{\lambda}\, s}}\, \Big[\pi_{\theta}\, {\hat b}_{\theta} +
  {{\pi_{\phi}}\over {\sin \theta}}\, {\hat b}_{\phi}\Big](\tau
 ,\lambda ).
 \label{5.11}
 \eeq

We have the following relation between $\pi_s$ and the Abelianized
constraint $\pi_{\chi}$

\bea
 \pi_{\chi}(\tau ,\lambda ) &=& \Big[\partial_{\lambda}\,
 {\vec {\cal P}} \cdot \hat t\Big](\tau ,\lambda ) = \Big[
 \pi_s - \partial_{\lambda}\, \theta\, {\vec {\cal P}} \cdot
 {\hat b}_{\theta} - \sin \theta\, \partial_{\lambda}\, \phi\,
 {\vec {\cal P}} \cdot {\hat b}_{\phi}\Big](\tau \lambda ) =\nonumber \\
 &&\nonumber\\
 &=& \Big[
 \pi_s + {{\partial_{\lambda}\, \theta\, \pi_{\theta} + \partial_{\lambda}\,
 \phi\, \pi_{\phi}}\over {\partial_{\lambda}\, s}}\Big](\tau ,\lambda ),\nonumber \\
 &&{}\nonumber \\
 &&\{ \theta (\tau ,\lambda ), \pi_{\chi}(\tau ,\lambda_1)\} =
  {{\partial_{\lambda}\, \theta}\over {\partial_{\lambda}\, s}}
 (\tau ,\lambda )\, \delta(\lambda - \lambda_1),\nonumber \\
 &&\nonumber\\
 &&\{ \phi (\tau ,\lambda ), \pi_{\chi}(\tau ,\lambda_1)\} =
  {{\partial_{\lambda}\, \phi}\over {\partial_{\lambda}\, s}}
 (\tau ,\lambda )\, \delta(\lambda - \lambda_1).
 \label{5.12}
 \eea

\bigskip

As a consequence, the functions $\theta (\tau ,\lambda )$ and $\phi
(\tau ,\lambda )$ are {\it not gauge invariant}. To find the two
corresponding Dirac observables we have to solve the following
multi-temporal equations (generated by the constraint $\chi (\tau
,\lambda ) \approx 0$), which determine the dependence of $\theta
(\tau ,\lambda )$ and $\phi (\tau ,\lambda )$ on the gauge variable
$s(\tau ,\lambda )$

\bea
 \partial_{\lambda}\, s(\tau ,\lambda )\,
 {{\delta\, \theta (\tau ,\lambda )}\over {\delta\, s(\tau ,\lambda_1)}}
 &{\buildrel {def}\over =}& \{ \theta (\tau ,\lambda ), \chi (\tau
 ,\lambda_1)\} = {{\partial\, \theta (\tau ,\lambda )}\over
 {\partial\, \lambda}}\, \delta (\lambda - \lambda_1),\nonumber \\
 &&\nonumber\\
  \partial_{\lambda}\, s(\tau ,\lambda )\,
 {{\delta\, \phi (\tau ,\lambda )}\over {\delta\, s(\tau ,\lambda_1)}}
 &{\buildrel {def}\over =}& \{ \phi (\tau ,\lambda ), \chi (\tau
 ,\lambda_1)\} = {{\partial\, \phi (\tau ,\lambda )}\over
 {\partial\, \lambda}}\, \delta (\lambda - \lambda_1).
 \label{5.13}
 \eea

\medskip

If we introduce the following ansatz

\bea
 F(\tau ,\lambda ) &=& \bar F(\tau ,s(\tau ,\lambda )) = T_{\lambda
 \mapsto s(\tau ,\lambda )}\, \bar F(\tau ,\lambda ),\qquad F =
 \theta ,\, \phi ,\nonumber \\
 &&{}\nonumber \\
 \Rightarrow &&
 \{ \bar \theta (\tau ,\lambda ), s (\tau ,\lambda_1)\} =
 \{ \bar \phi (\tau ,\lambda ), s (\tau ,\lambda_1)\} = 0,
 \nonumber  \\
 &&{}\nonumber \\
 T_{\lambda \mapsto s(\tau ,\lambda )} &=& \sum_o^{\infty}\,
 {1\over {n!}}\, \Big(s(\tau ,\lambda ) - \lambda \Big)^n\,
 {{\partial^n}\over {\partial \lambda^n}},
 \label{5.14}
 \eea

\noindent and we remember the results $\chi(\tau,\lambda) =
\partial_\lambda s(\tau,\lambda)\,\pi_\chi(\tau,\lambda)$ and
$\{s(\tau,\lambda),\chi(\tau,\lambda_1)\}= \partial_\lambda
s(\tau,\lambda)\,\delta(\lambda-\lambda_1)$, then we have

\bea
 && \partial_{\lambda}\, s(\tau ,\lambda )\,
 {{\delta\, \bar \theta (\tau ,s(\tau ,\lambda ))}\over {\delta\, s(\tau ,\lambda_1)}}
 = \{ \bar \theta(\tau ,s(\tau ,\lambda )), \chi (\tau
 ,\lambda_1)\} =\nonumber \\
 &&\nonumber\\
 &=& \partial_\lambda s(\tau,\lambda)\,\delta (\lambda - \lambda_1)\, T_{\lambda \mapsto s(\tau ,\lambda
 )}\, {{\partial\, \bar \theta (\tau ,\lambda )}\over {\partial\,
 \lambda }} + T_{\lambda \mapsto s(\tau ,\lambda )}\, \{ \bar
 \theta (\tau ,\lambda ), \chi (\tau ,\lambda_1)\} =\nonumber \\
 &&\nonumber\\
 &=& {{\partial\, \theta (\tau ,\lambda )}\over
 {\partial\, \lambda}}\, \delta (\lambda - \lambda_1) +
 T_{\lambda \mapsto s(\tau ,\lambda )}\, \{ \bar
 \theta (\tau ,\lambda ), \chi (\tau ,\lambda_1)\},\nonumber \\
 &&\nonumber\\
 &&\nonumber\\
 &&\partial_{\lambda}\, s(\tau ,\lambda )\,
 {{\delta\, \bar \phi(\tau ,s(\tau ,\lambda ))}\over {\delta\,
 s(\tau ,\lambda_1)}}
 = \{ \bar \phi (\tau ,s(\tau ,\lambda )), \chi (\tau
 ,\lambda_1)\} =\nonumber \\
 &&\nonumber\\
 &=& \partial_\lambda s(\tau,\lambda)\,\delta (\lambda - \lambda_1)\, T_{\lambda \mapsto s(\tau ,\lambda
 )}\, {{\partial\, \bar \phi (\tau ,\lambda )}\over {\partial\,
 \lambda }} + T_{\lambda \mapsto s(\tau ,\lambda )}\, \{ \bar
 \phi (\tau ,\lambda ), \chi (\tau ,\lambda_1)\} =\nonumber \\
 &&\nonumber\\
 &=& {{\partial\, \phi (\tau ,\lambda )}\over
 {\partial\, \lambda}}\, \delta (\lambda - \lambda_1) +
 T_{\lambda \mapsto s(\tau ,\lambda )}\, \{ \bar
 \phi (\tau ,\lambda ), \chi (\tau ,\lambda_1)\},\nonumber \\
 &&{}\nonumber \\
 &&\Downarrow\nonumber \\
 &&{}\nonumber \\
 &&\{ \bar \theta (\tau ,\lambda ), \chi (\tau ,\lambda_1)\} =
 \{ \bar \phi (\tau ,\lambda ), \chi (\tau ,\lambda_1)\} = 0.
 \label{5.15}
 \eea
\medskip

Therefore {\it a possible set of configurational Dirac observables
for the open Nambu string} is

\bea
 \bar \theta (\tau ,\lambda ) &=& T^{-1}_{\lambda \mapsto s(\tau
,\lambda )}\, \theta (\tau ,\lambda ),\nonumber \\
&&\nonumber\\
 \bar \phi(\tau ,\lambda ) &=& T^{-1}_{\lambda \mapsto s(\tau ,\lambda )}\,
 \phi(\tau ,\lambda ).
 \label{5.16}
 \eea

\bigskip

As a consequence, after the canonical transformation (\ref{5.7}) we
must do another Shanmugadhasan point canonical transformation
leading to a canonical basis adapted to the Abelianized constrains
$\pi_{\chi}(\tau, \lambda) \approx 0$ and containing a canonical
basis of Dirac observables for the open string

\bea
 &&\begin{minipage}[t]{4 cm}
\begin{tabular}{|l|l|l|} \hline
$s(\tau ,\lambda)$ & $\theta (\tau ,\lambda )$ & $\phi (\tau ,\lambda )$\\
\hline  $\pi_s(\tau ,\lambda )$ &
$\pi_{\theta}(\tau ,\lambda )$ & $\pi_{\phi}(\tau ,\lambda )$\\
\hline
\end{tabular}
\end{minipage} \hspace{1cm} {\longrightarrow \hspace{.2cm}} \
\begin{minipage}[t]{4 cm}
\begin{tabular}{|l|l|l|} \hline
$s(\tau ,\lambda)$ & $\bar \theta (\tau ,\lambda )$ & $\bar \phi (\tau ,\lambda )$\\
\hline  $\pi_{\chi}(\tau ,\lambda ) \approx 0$ &
$\pi_{\bar \theta}(\tau ,\lambda )$ & $\pi_{\bar \phi}(\tau ,\lambda )$\\
\hline
\end{tabular}
\end{minipage},\nonumber \\
 &&{}\nonumber \\
 &&{}\nonumber \\
 &&\{ s(\tau ,\lambda ), \pi_{\chi}(\tau ,\lambda_1)\} = \{ \bar \theta
 (\tau ,\lambda ), \pi_{\bar \theta}(\tau ,\lambda_1)\} = \{ \bar \phi (\tau
 ,\lambda ), \pi_{\bar \phi}(\tau ,\lambda_1)\} = \delta(\lambda
-\lambda_1).
 \label{5.17}
 \eea
\medskip

The generating function of the point Shanmugadhasan canonical
transformation (\ref{5.17}) is

\bea
 \Phi_1 &=& \int d\lambda \Big[\pi_s\, s + \pi_{\theta}\,
 T_{\lambda \mapsto s(\tau ,\lambda )}\, \bar \theta
 + \pi_{\phi}\, T_{\lambda \mapsto s(\tau ,\lambda )}\,
 \bar \phi \Big](\tau ,\lambda ) =\nonumber \\
 &=& \int d\lambda\, \Big[\Big(\pi_s\, s\Big)(\tau ,\lambda ) +
 \pi_{\theta}(\tau ,\lambda )\, \bar \theta (\tau ,s(\tau ,\lambda)) +
 \pi_{\phi}(\tau ,\lambda )\, \bar \phi (\tau ,s(\tau ,\lambda ))\Big],\nonumber \\
 &&{}\nonumber \\
  \pi_{\chi}(\tau ,\lambda ) &=& {{\delta \Phi_1}\over {\delta s(\tau
  ,\lambda )}},\nonumber \\
    \pi_{\bar \theta}(\tau ,\lambda ) &=& {{\delta \Phi_1}\over {\delta \bar{\theta} (\tau
  ,\lambda )}}, \nonumber \\
    \pi_{\bar \phi}(\tau ,\lambda ) &=& {{\delta \Phi_1}\over {\delta \bar{\phi} (\tau
  ,\lambda )}}.
 \label{5.18}
 \eea
\medskip

\medskip

The explicit expression of the new momenta is \footnote{ We use
$\theta (\tau ,\lambda ) = \bar \theta (\tau ,s(\tau ,\lambda )),
\qquad\phi (\tau ,\lambda ) = \bar \phi (\tau ,s(\tau ,\lambda ))$
and then we have $\hat t(\tau \lambda ) = {\hat {\bar t}}(\tau
,s(\tau ,\lambda )), \qquad{\hat b}_{\theta}(\tau \lambda ) = {\hat
{\bar b}}_{\bar \theta}(\tau ,s(\tau ,\lambda )), \qquad{\hat
b}_{\phi}(\tau \lambda ) = {\hat {\bar b}}_{\bar \phi}(\tau ,s(\tau
,\lambda ))$. We use $\partial_s\, \bar \theta (\tau ,s(\tau
,\lambda)) =
\partial_{\lambda}\, \theta (\tau ,\lambda )/\partial_{\lambda}\,
s(\tau ,\lambda )$, the analogous formula for $\bar \phi$ and
$\delta (\lambda - s(\tau ,\lambda_1)) = \delta (\lambda_1
-s^{-1}(\tau ,\lambda )) / |\partial_{\lambda_1}\, s(\tau
,\lambda_1)|_{\lambda_1 = s^{-1}(\tau ,\lambda )}$. }

\begin{eqnarray*}
 \pi_{\chi}(\tau ,\lambda ) &=& \pi_s(\tau ,\lambda ) +
 \pi_{\theta}(\tau ,\lambda )\, \partial_s\, \bar \theta (\tau
 ,s(\tau ,\lambda )) + \pi_{\phi}(\tau ,\lambda )\, \partial_s\,
 \bar \phi (\tau ,s(\tau ,\lambda )) =\nonumber \\
 &&\nonumber\\
 &=& \Big[\pi_s + {{\partial_{\lambda}\, \theta\, \pi_{\theta} +
 \partial_{\lambda}\, \phi\, \pi_{\phi}}\over {\partial_{\lambda}\, s}}
  \Big](\tau ,\lambda ) \approx 0,\nonumber \\
  &&\nonumber\\
  &&\nonumber\\
 \pi_{\bar \theta}(\tau ,\lambda ) &=& \int d\lambda_1\,
 \pi_{\theta}(\tau ,\lambda_1)\, T_{\lambda_1 \mapsto s(\tau ,\lambda_1 )}\,
 \delta (\lambda - \lambda_1) =\nonumber \\
 &&\nonumber\\
 &=& \int d\lambda_1\, \pi_{\theta}(\tau ,\lambda_1)\, \delta
 (\lambda - s(\tau ,\lambda_1)) = \Big(\partial_{\bar \lambda}\, s(\tau ,\bar \lambda )
 |_{\bar \lambda = s^{-1}(\tau ,\lambda )}\Big)^{-1}\,
 \pi_{\theta}(\tau ,s^{-1}(\tau ,\lambda)),\nonumber \\
 &&\nonumber\\
 &&\nonumber\\
 \pi_{\bar \phi} (\tau ,\lambda )&=& \int d\lambda_1\,
 \pi_{\phi}(\tau ,\lambda_1)\, T_{\lambda_1 \mapsto s(\tau ,\lambda_1 )}\,
 \delta (\lambda - \lambda_1) =\nonumber \\
 &&\nonumber\\
 &=& \int d\lambda_1\, \pi_{\phi}(\tau ,\lambda_1)\, \delta
 (\lambda - s(\tau ,\lambda_1)) = \Big(\partial_{\bar \lambda}\, s(\tau ,\bar \lambda )
 |_{\bar \lambda = s^{-1}(\tau ,\lambda )}\Big)^{-1}\,
 \pi_{\phi}(\tau ,s^{-1}(\tau ,\lambda)),\nonumber \\
  &&{}\nonumber \\
  &&\Downarrow
  \end{eqnarray*}

\bea
 \pi_{\theta}(\tau ,\lambda ) &=& \partial_{\lambda}\, s(\tau
 ,\lambda )\, \pi_{\bar \theta}(\tau ,s(\tau ,\lambda )),\nonumber \\
 &&\nonumber\\
 \pi_{\phi}(\tau ,\lambda ) &=& \partial_{\lambda}\, s(\tau
 ,\lambda )\, \pi_{\bar \phi}(\tau ,s(\tau ,\lambda )),\nonumber \\
 &&\nonumber\\
 \pi_s(\tau ,\lambda ) &=& \pi_{\chi}(\tau ,\lambda ) -
 \partial_{\lambda}\, s(\tau ,\lambda )\, \Big[\partial_s\,
 \bar \theta\, \pi_{\bar \theta} + \partial_s\, \bar \phi\,
 \pi_{\bar \phi}\Big](\tau ,s(\tau ,\lambda )) =\nonumber \\
 &&\nonumber\\
 &=& \pi_{\chi}(\tau ,\lambda ) - \Big[\partial_{\lambda}\, \bar \theta\,
 \pi_{\bar \theta} + \partial_{\lambda}\, \bar \phi\, \pi_{\bar \phi}
 \Big](\tau ,s(\tau ,\lambda )),\nonumber \\
 &&{}\nonumber \\
 &&{}\nonumber \\
 \vec y(\tau ,\lambda ) &=& - \partial_{\lambda}\, s(\tau ,\lambda
 )\, {\hat {\bar t}}(\tau ,s(\tau ,\lambda )),\nonumber \\
 &&\nonumber\\
 {\vec {\cal P}}(\tau ,\lambda ) &=& {\hat {\bar
 t}}(\tau ,s(\tau ,\lambda ))\,\int_o^{\lambda} d\lambda_1\,\Big[
 \pi_{\chi}(\tau ,\lambda_1) - \Big(\partial_{\lambda_1}\, \bar \theta\,
 \pi_{\bar \theta} + \partial_{\lambda_1}\, \bar \phi\, \pi_{\bar \phi}
 \Big)(\tau ,s(\tau ,\lambda_1 ))\Big] -\nonumber \\
 &&\nonumber\\
 &-& \Big[\pi_{\bar \theta}\, {\hat {\bar b}}_{\bar \theta} +
 {{\pi_{\bar \phi}}\over {\sin \bar \theta}}\, {\hat {\bar b}}_{\bar
 \phi}\Big](\tau ,s(\tau ,\lambda )).
 \label{5.19}
 \eea

\medskip

The invariant mass becomes

\bea
 &&M = \int_o^{\pi} d\lambda\, \sqrt{-N^2\, \Big(\partial_{\lambda}\, s(\tau
 ,\lambda)\Big)^2 + (\partial_\lambda \vec {\cal P})^2(\tau
 ,\lambda  )},\nonumber \\
 &&{}\nonumber \\
 &&\nonumber\\
 &&\partial_\lambda \vec {\cal P}(\tau ,\lambda ) \approx
 {\hat {\bar t}}(\tau ,s(\tau ,\lambda ))\,\pi_{\chi}(\tau ,\lambda) +\nonumber\\
 &&\nonumber\\
 &+&\hat{\bar{b}}_{\bar{\theta}}\,
 \Big[
 -\partial_\lambda\bar{\theta}\,\int_o^{\lambda} d\lambda_1\,\Big(\partial_{\lambda_1}\, \bar \theta\,
 \pi_{\bar \theta} + \partial_{\lambda_1}\, \bar \phi\, \pi_{\bar \phi}
 \Big)(\tau ,s(\tau ,\lambda_1 ))\nonumber\\
 &&\nonumber\\
 &&-\partial_\lambda\pi_{\bar{\theta}}+
 \frac{\cos\bar{\theta}}{\sin\bar{\theta}}\,\pi_{\bar{\phi}}\,\partial_\lambda\bar{\phi}
 \Big](\tau ,s(\tau ,\lambda ))+\nonumber\\
 &&\nonumber\\
 &+&\hat{\bar{b}}_{\bar{\phi}}\,
 \Big[-\sin\bar{\theta}\,\partial_\lambda\bar{\phi}\,\int_o^{\lambda} d\lambda_1\,\Big(\partial_{\lambda_1}\, \bar \theta\,
 \pi_{\bar \theta} + \partial_{\lambda_1}\, \bar \phi\, \pi_{\bar \phi}
 \Big)(\tau ,s(\tau ,\lambda_1 ))\nonumber\\
 &&\nonumber\\
 && -\frac{\partial_\lambda\pi_{\bar{\phi}}}{\sin\bar{\theta}}+
 \frac{\cos\bar{\theta}}{\sin2\bar{\theta}}\,\pi_{\bar{\phi}}\,\partial_\lambda\bar{\theta}-
 \cos\bar{\theta}\,\pi_{\bar{\theta}}\,\partial_\lambda\bar{\phi}
 \Big](\tau ,s(\tau ,\lambda ))
 \label{5.20}
 \eea

\bigskip

We can now restrict ourselves to the FS gauge $s(\tau, \lambda)
\approx \lambda$, which identifies the parameter $\lambda$ with the
arc-length along the open string. In this way the open string is
completely described in terms of the external center of mass (the
frozen Jacobi data $\vec z$ and $\vec h$) and of the Dirac
observables $\bar \theta$, $\pi_{\bar \theta}$, $\bar \phi$,
$\pi_{\bar \phi}$. In the FS gauge we get (the Fokker-Pryce center
of inertia $Y^{\mu}(\tau)$ is given in Eq.(\ref{4.2}))

 \begin{eqnarray*}
  x^{\mu}(\tau ,\lambda ) &=& Y^{\mu}(\tau) +
 \epsilon^{\mu}_r(\vec h)\, \eta^r(\tau ,\lambda ),\nonumber \\
 &&{}\nonumber \\
   \vec \eta (\tau ) &\approx& +
{1\over { 2\pi\,\int_o^{\pi} d\lambda \, \sqrt{-N^2\,   + \Big(
{\partial_\lambda \vec {\cal P}}(\tau ,\lambda )\Big)^2}
 }} \nonumber \\
 &&  \int_o^{\pi} d\lambda\,
 \int^{\pi}_{-\pi} d\lambda_1\, \int_\lambda^{\lambda_1}\,
 \vec y(\tau ,\lambda_2)\,
 \sqrt{-N^2\,   + \Big( {\partial_\lambda \vec {\cal P}}(\tau ,\lambda )\Big)^2},\nonumber \\
 &&{}\nonumber \\
 \vec y(\tau ,\lambda ) &=& - {\hat {\bar t}}(\tau ,\lambda
 ) = - \Big(\sin \bar \theta\, \cos \bar \phi, \sin \bar \theta\,
 \sin \bar \phi, \cos \bar \theta\Big)(\tau, \lambda),\nonumber \\
 {\vec {\cal P}}(\tau ,\lambda ) &=& - \int_o^{\lambda} d\lambda_1\,\Big[
   \Big(\partial_{\lambda_1}\, \bar \theta\,
 \pi_{\bar \theta} + \partial_{\lambda_1}\, \bar \phi\, \pi_{\bar \phi}
 \Big)(\tau ,\lambda_1)\Big]\, {\hat {\bar
 t}}(\tau ,\lambda ) - \Big[\pi_{\bar \theta}\, {\hat {\bar b}}_{\bar \theta} +
 {{\pi_{\bar \phi}}\over {\sin \bar \theta}}\, {\hat {\bar b}}_{\bar
 \phi}\Big](\tau ,\lambda ),
 \end{eqnarray*}

\bea
 M &=& \int_o^{\pi} d\lambda\, \sqrt{-N^2 + A^2(\tau ,\lambda )+B^2(\tau ,\lambda )},\nonumber \\
 &&{}\nonumber \\
 {\partial_\lambda \vec {\cal P}}(\tau ,\lambda ) &=&
 \hat{\bar{b}}_{\bar{\theta}}(\tau ,\lambda )\,A(\tau,\lambda)
 +\hat{\bar{b}}_{\bar{\phi}}(\tau ,\lambda )\,B(\tau,\lambda),\nonumber \\
 &&{}\nonumber \\
 A(\tau,\lambda)&=& -\partial_\lambda\bar{\theta}(\tau ,\lambda )\,
 \int_o^{\lambda} d\lambda_1\,\Big(\partial_{\lambda_1}\, \bar \theta\,
 \pi_{\bar \theta} + \partial_{\lambda_1}\, \bar \phi\, \pi_{\bar \phi}
 \Big)(\tau ,\lambda_1) -\nonumber\\
 &&\nonumber\\
 &&-\partial_\lambda\pi_{\bar{\theta}}(\tau ,\lambda ) +
 \Big[\frac{\cos\bar{\theta}}{\sin\bar{\theta}}\,\pi_{\bar{\phi}}\,\partial_\lambda\bar{\phi}
 \Big](\tau ,\lambda ),\nonumber \\
 &&{}\nonumber \\
 B(\tau,\lambda)&=& - \Big[\sin\bar{\theta}\,\partial_\lambda\bar{\phi}\Big](\tau ,\lambda )\,
 \int_o^{\lambda} d\lambda_1\,\Big(\partial_{\lambda_1}\, \bar \theta\,
 \pi_{\bar \theta} + \partial_{\lambda_1}\, \bar \phi\, \pi_{\bar \phi}
 \Big)(\tau ,\lambda_1) -\nonumber\\
 &&\nonumber\\
 && -\frac{\partial_\lambda\pi_{\bar{\phi}}(\tau ,\lambda )}{\sin\bar{\theta}(\tau ,\lambda )}+
 \Big[\frac{\cos\bar{\theta}}{\sin2\bar{\theta}}\,\pi_{\bar{\phi}}\,\partial_\lambda\bar{\theta}-
 \cos\bar{\theta}\,\pi_{\bar{\theta}}\,\partial_\lambda\bar{\phi}
 \Big](\tau ,\lambda)
 ,\nonumber \\
 &&\nonumber\\
 &&{}\nonumber \\
 \vec S &=& \int^{\pi}_{o} d\lambda\,
 \Big[+ \pi_{\bar \theta}\, {\hat {\bar b}}_{\bar \phi} -
 {{\pi_{\bar \phi}}\over {\sin \bar \theta}}\,
 {\hat {\bar b}}_{\bar \theta}\Big](\tau ,\lambda ),\nonumber \\
 &&{}\nonumber \\
  \rho_1(\tau ,\lambda ) &=&  \sqrt{(\partial_{\lambda}\, \bar \theta)^2 +
 \sin^2\, \bar \theta\, (\partial_{\lambda}\, \bar \phi)^2}(\tau ,\lambda
 ),\nonumber \\
 &&{}\nonumber \\
 \rho_2(\tau ,\lambda ) &=& {{\sin \bar \theta\, (\partial_{\lambda}\,
 \bar \phi\, \partial^2_{\lambda}\, \bar \theta - \partial_{\lambda}\,
 \bar \theta\, \partial^2_{\lambda}\, \bar \phi) - \sin2\, \bar \theta\,
 \cos \bar \theta\, (\partial_{\lambda}\, \bar \phi)^3 - 2\, \cos \bar \theta\,
 (\partial_{\lambda}\, \bar \theta )^2\, \partial_{\lambda}\, \bar \phi }\over
 { \Big[(\partial_{\lambda}\, \bar \theta)^2 + \sin^2\, \bar \theta\,
 (\partial_{\lambda}\, \bar \phi)^2\Big]}}(\tau ,\lambda ).\nonumber \\
 &&{}
 \label{5.21}
 \eea

\noindent where Eq.(\ref{4.9}) was used.

\medskip

The resulting Hamilton equations for the Dirac observables must be
integrated with the boundary conditions ${\dot {\vec \eta}}^2_o(\tau
) = {\dot {\vec \eta}}^2_{\pi}(\tau ) = 1$ (${\dot {\vec
\eta}}_A(\tau ) = \partial_{\tau}\, \vec \eta (\tau ,\lambda
){|}_{\lambda = A}$, $A= 0, \pi$), which imply that the end points
of the string ${\vec \eta}_A(\tau ) = \vec \eta (\tau ,A)$, $A = 0,
\pi$, move on null curves. Note that we have ${\vec \eta}_o(\tau ) -
{\vec \eta}_{\pi}(\tau ) = \int_o^{\pi} d\lambda\, {\hat {\bar
t}}(\tau ,\lambda )$. Moreover we have ${\hat {\bar t}}_A(\tau ) =
{\hat {\bar t}}(\tau ,A) = - \vec y(\tau ,A)$, $A = 0, \pi$, with
${\hat {\bar t}}_{\pi}(\tau ) - {\hat {\bar t}}_o(\tau ) =
\int_o^{\pi} d\lambda\,
\partial_{\lambda}\, {\hat {\bar t}}(\tau ,\lambda )$.

\bigskip

The Hamilton equations generated by the Hamiltonian $M$ are

\bea
 \partial_{\tau}\, \bar \theta (\tau ,\lambda ) &=&
 -\partial_\lambda{\bar\theta}(\tau ,\lambda )\int_\lambda^\pi
 d\lambda'\,\left[
 \frac{\partial_\lambda{\bar\theta}\,A+\sin{\bar\theta}\,\partial_\lambda{\bar\phi}\,B}
 { \sqrt{- N^2 + A^2 + B^2}}
 \right](\tau,\lambda')+\nonumber\\
 &&\nonumber\\
 &+&\partial_\lambda\left(
 \frac{A}{\sqrt{- N^2 + A^2 + B^2}}
 \right)(\tau,\lambda)
 -\left(\frac{\cos{\bar\theta}\,\partial_\lambda{\bar\phi}\,B}{\sqrt{- N^2 + A^2 + B^2}}
 \right)(\tau,\lambda)
 ,\nonumber \\
 &&\nonumber\\
 &&\nonumber\\
 \partial_{\tau}\, \bar \phi (\tau ,\lambda ) &=& -
 \partial_\lambda{\bar\phi}(\tau ,\lambda )\int_\lambda^\pi
 d\lambda'\,\left[
 \frac{\partial_\lambda{\bar\theta}\,A+\sin{\bar\theta}\,\partial_\lambda{\bar\phi}\,B}
 { \sqrt{- N^2 + A^2 + B^2}}
 \right](\tau,\lambda')+\nonumber\\
 &&\nonumber\\
 &+&\partial_\lambda\left(
 \frac{B}{\sin{\bar\theta}\,\sqrt{- N^2 + A^2 + B^2}}
 \right)(\tau,\lambda)
 +\left(\frac{
 \frac{\cos{\bar\theta}}{\sin{\bar\theta}}\,\partial_\lambda{\bar\phi}\,A-
 \frac{\cos{\bar\theta}}{\sin2{\bar\theta}}\,\partial_\lambda{\bar\theta}\,B}
 {\sqrt{- N^2 + A^2 + B^2}}
 \right)(\tau,\lambda),\nonumber \\
 &&{}\nonumber \\
 &&{}\nonumber \\
 \partial_{\tau}\, \pi_{\bar \theta}(\tau ,\lambda ) &=&
 -\partial_\lambda\pi_{\bar \theta}(\tau ,\lambda )\int_\lambda^\pi
 d\lambda'\,\left[
 \frac{\partial_\lambda{\bar\theta}\,A+\sin{\bar\theta}\,\partial_\lambda{\bar\phi}\,B}
 { \sqrt{- N^2 + A^2 + B^2}}
 \right](\tau,\lambda')+\nonumber\\
 &&\nonumber\\
 &+&\partial_\lambda\left(
 \frac{\frac{\cos{\bar\theta}}{\sin{\bar\theta}\,}\pi_{\bar\phi}\,B}{\sqrt{- N^2 + A^2 + B^2}}
 \right)(\tau,\lambda)+\nonumber\\
 &&\nonumber\\
 &-&\left(
 \frac{
 \Big(\pi_{\bar\theta}\,\partial_\lambda{\bar\theta}+\frac{{\bar\phi}\,
 \partial_\lambda{\bar\phi}}{\sin2{\bar\theta}}
 \Big)A+ \Big(
\frac{\cos{\bar\theta}}{\sin{\bar\theta}}\,\partial_\lambda\pi_{\bar\phi}+
\frac{1+\cos2{\bar\theta}}{\sin2{\bar\theta}}\,\pi_{\bar\phi}\,\partial_\lambda
 {\bar\theta} \Big)\,B }{\sqrt{- N^2 + A^2 + B^2}} \right)(\tau,\lambda)
 ,\nonumber \\
 &&\nonumber\\
 &&\nonumber\\
 \partial_{\tau}\, \pi_{\bar \phi}(\tau ,\lambda ) &=&
 -\partial_\lambda\pi_{\bar \phi}(\tau ,\lambda )\int_\lambda^\pi
 d\lambda'\,\left[
 \frac{\partial_\lambda{\bar\theta}\,A+\sin{\bar\theta}\,\partial_\lambda{\bar\phi}\,B}
 { \sqrt{- N^2 + A^2 + B^2}}
 \right](\tau,\lambda')+\nonumber\\
 &&\nonumber\\
 &+&\partial_\lambda\left(
 \frac{\frac{\cos{\bar\theta}}{\sin{\bar\theta}\,}\pi_{\bar\phi}\,A+
 \cos{\bar\theta}\,\pi_{\bar\theta}\,B}{\sqrt{- N^2 + A^2 + B^2}}
 \right)(\tau,\lambda)+\nonumber\\
 &&\nonumber\\
 &-&\left(\frac{
 \pi_{\bar\phi}\,\partial_\lambda{\bar\theta}\,A+\sin{\bar\theta}\,
 \partial_\lambda{\bar\phi}\,\pi_{\bar\phi}\,B
 }{\sqrt{- N^2 + A^2 + B^2}}
 \right)(\tau,\lambda).
 \label{5.22}
 \eea

\bigskip

This series of  {\it point}  canonical transformations should be
completed with the following one

\bea
 &&\begin{minipage}[t]{3cm}
\begin{tabular}{|l|} \hline
$\vec y (\tau ,\lambda )$ \\ \hline ${\vec {\cal P}}(\tau ,\lambda
)$\\ \hline
\end{tabular}
\end{minipage} \hspace{1cm} {\longrightarrow \hspace{.2cm}} \
\begin{minipage}[t]{4 cm}
\begin{tabular}{|l|l|l|} \hline
$s(\tau ,\lambda)$ & $\bar \theta(\tau ,\lambda )$ & $\bar \phi (\tau ,\lambda )$\\
\hline  $\pi_{\chi}(\tau ,\lambda ) \approx 0$ &
$\pi_{\bar \theta}(\tau ,\lambda )$ & $\pi_{\bar \phi}(\tau ,\lambda )$\\
\hline
\end{tabular}
\end{minipage}\nonumber \\
 &&{}\nonumber \\
 &&{}\nonumber \\
 &&\hspace{1cm} {\longrightarrow \hspace{.2cm}} \
\begin{minipage}[t]{4 cm}
\begin{tabular}{|l|l|l|} \hline
$s(\tau ,\lambda)$ & $\rho_1(\tau ,\lambda )$ & $\rho_2(\tau ,\lambda )$\\
\hline  $\pi_{\chi}(\tau ,\lambda ) \approx 0$ &
$\pi_1(\tau ,\lambda )$ & $\pi_2(\tau ,\lambda )$\\
\hline
\end{tabular}
\end{minipage}
 \label{5.23}
 \eea
\medskip

\noindent in which the Dirac observables $\bar \theta$, $\pi_{\bar
\theta}$, $\bar \phi$, $\pi_{\bar \phi}$, are replaced with the
Lorentz scalar Dirac observables $\rho_1$, $\pi_1$, $\rho_2$,
$\pi_2$ given in Eqs.(\ref{5.21}). Since we have $\rho_1(\tau
,\lambda ) \geq 0$ and $\rho_2(\tau ,\lambda ) \geq 0$, these
variables are like radial variables \footnote{In trying to quantize
them one should use the non-canonical basis $r$ and $r\, p_r$, the
only one which would give meaningful commutators due to $r \geq
0$.}.

\bigskip

Regarding the last canonical transformation (\ref{5.23}), we know
${\bar \rho}_i = {\bar \rho}_i(\bar \theta , \bar \phi)$, $i=1,2$,
from Eqs.(\ref{5.21}), but not the inversion $\bar \theta = \bar
\theta ({\bar \rho}_i)$, $\bar \phi = \bar \phi ({\bar \rho}_i)$.
Therefore we can only define the generating function of the inverse
canonical transformation

\bea
 \Psi &=& \int d\lambda\, \Big[{\bar \pi}_1\, {\bar \rho}_1(\bar \theta ,\bar \phi)
 + {\bar \pi}_2\, {\bar \rho}_2(\bar \theta ,\bar \phi)\Big](\tau ,\lambda
 ), \nonumber \\
 &&{}\nonumber \\
 \pi_{\bar \theta}(\tau ,\lambda ) &=& {{\delta \Psi}\over {\delta\,
 \bar \theta (\tau ,\lambda )}},\nonumber \\
  \pi_{\bar \phi}(\tau ,\lambda ) &=& {{\delta \Psi}\over {\delta\,
 \bar \phi (\tau ,\lambda )}}.
 \label{5.24}
 \eea

It can give  ${\bar \pi}_i = {\bar \pi}_i(\bar \theta, \pi_{\bar
\theta}, \bar \phi , \pi_{\bar \phi})$, linear in $\pi_{\bar
\theta}$ and $\pi_{\bar \phi}$.\medskip

However, to find the expression of the invariant mass $M$ we need
the inversion  $\bar \theta ({\bar \rho}_i)$, $\bar \phi ({\bar
\rho}_i)$, $\pi_{\bar \theta} ({\bar \rho}_i, {\bar \pi}_i)$,
$\pi_{\bar \phi} ({\bar \rho}_i, {\bar \pi}_i)$ \footnote{The
existence of the inversion is assured by theorem (7.5) in  chapter 7
of Ref.\cite{23a}. It implies that, according to the fundamental
theorem for curves, the embedding functions $\vec \eta (\tau
,\lambda )$, and therefore $\vec y (\tau ,\lambda ) = {\hat {\bar
t}}(\tau ,\lambda ) = - \partial_{\lambda }\, \vec \eta (\tau
,\lambda )$, can be obtained from the Frenet-Serret curvatures
${\bar \rho}_i(\tau ,\lambda )$ by quadratures.}.

\vfill\eject

\section{Particle-like Variables for the Transverse Oscillations
and the Problem of Quantization.}

Since we cannot find explicitly the canonical transformation to the
canonical basis of Dirac observables $\rho_i$, $\pi_i$, whose
inverse is defined in Eq.(\ref{5.23}), let us look for a different
canonical basis of Dirac observables, which avoids the angles
because they cannot be quantized consistently.\medskip

If we interpret the angles and their conjugate momenta as
angle-action-like variables, the following canonical transformation
emerges in a natural way [$N/c$ has dimensions $m\, t^{-1}$ like
$m\, \omega$ in the harmonic oscillator $H = {{p^2}\over {2m}} +
{1\over 2}\, m\, \omega^2\, q^2 = \omega\, I$, with $I$ action
variable]

\bea
 q_{\bar \theta}(\tau ,\lambda ) &=& +\sqrt{{{2c}\over N}\, \pi_{\bar \theta}
 (\tau ,\lambda )}\, \cos \bar \theta (\tau ,\lambda ),\nonumber \\
 p_{\bar \theta}(\tau ,\lambda ) &=& - \sqrt{{{2N}\over c}\, \pi_{\bar \theta}
 (\tau ,\lambda )}\, \sin \bar \theta (\tau ,\lambda ),\nonumber \\
 q_{\bar \phi}(\tau ,\lambda ) &=& +\sqrt{{{2c}\over N}\, \pi_{\bar \phi}
 (\tau ,\lambda )}\, \cos \bar \phi (\tau ,\lambda ),\nonumber \\
 p_{\bar \phi}(\tau ,\lambda ) &=& -\sqrt{{{2N}\over c}\, \pi_{\bar \phi}
 (\tau ,\lambda )}\, \sin \bar \phi (\tau ,\lambda ),\nonumber \\
 &&{}\nonumber \\
 &&\{ q_{\bar \theta}(\tau ,\lambda ), p_{\bar \theta}(\tau
 ,\lambda_1)\} = \{ q_{\bar \phi}(\tau ,\lambda ), p_{\bar \phi}(\tau
 ,\lambda_1)\} = \delta (\lambda - \lambda_1),\nonumber \\
 &&{}\nonumber \\
 &&\sin \bar \theta (\tau ,\lambda ) = - {{p_{\bar \theta}}\over
 {\sqrt{\frac{N^2}{c^2}\,q^2_{\bar \theta} + p^2_{\bar \theta}}}}(\tau
 ,\lambda ),\qquad \cos \bar \theta (\tau ,\lambda ) = {{q_{\bar \theta}}\over
 {\sqrt{q^2_{\bar \theta} + \frac{c^2}{N^2}\,p^2_{\bar \theta}}}}(\tau
 ,\lambda ),\nonumber \\
 &&\pi_{\bar \theta}(\tau ,\lambda ) = \left(\frac{N}{2c}\,q^2_{\bar \theta} +
 \frac{c}{2N}\,p^2_{\bar \theta}\right)(\tau ,\lambda ),\nonumber \\
 &&\nonumber\\
 &&\sin \bar \phi (\tau ,\lambda ) = - {{p_{\bar \phi}}\over
 {\sqrt{\frac{N^2}{c^2}\,q^2_{\bar \phi} + p^2_{\bar \phi}}}}(\tau
 ,\lambda ),\qquad \cos \bar \phi (\tau ,\lambda ) = {{q_{\bar \phi}}\over
 {\sqrt{q^2_{\bar \phi} + \frac{c^2}{N^2}\,p^2_{\bar \phi}}}}(\tau
 ,\lambda ),\nonumber \\
 &&\pi_{\bar \phi}(\tau ,\lambda ) = \left(\frac{N}{2c}\,q^2_{\bar \phi} +
 \frac{c}{2N}\,p^2_{\bar \phi}\right)(\tau ,\lambda ).
 \label{6.1}
 \eea

The particle-like variables $q_{\bar \theta}(\tau ,\lambda )$,
$p_{\bar \theta}(\tau ,\lambda )$, $q_{\bar \phi}(\tau ,\lambda )$,
$p_{\bar \phi}(\tau ,\lambda )$ describe the transverse oscillations
of the Nambu string.

\bigskip

We get \footnote{We use
\[
\partial_\lambda\sin{\bar\theta}(\tau ,\lambda ) = \cos{\bar\theta}(\tau ,\lambda )\,
\partial_\lambda{\bar\theta}(\tau ,\lambda )\,\Rightarrow\,
\partial_\lambda{\bar\theta}(\tau ,\lambda ) = \frac{\partial_\lambda\sin{\bar\theta}(\tau ,\lambda )}
{\cos{\bar\theta}(\tau ,\lambda )}
\]
and then we make the substitutions:
\[
\sin \bar \theta (\tau ,\lambda ) = - {{p_{\bar \theta}}\over
 {\sqrt{\frac{N^2}{c^2}\,q^2_{\bar \theta} + p^2_{\bar \theta}}}}(\tau
 ,\lambda ),\qquad \cos \bar \theta (\tau ,\lambda ) = {{q_{\bar \theta}}\over
 {\sqrt{q^2_{\bar \theta} + \frac{c^2}{N^2}\,p^2_{\bar \theta}}}}(\tau
 ,\lambda ),
\]
}

\bea
 \partial_{\lambda}\, \bar \theta(\tau ,\lambda ) &=&
\frac{p_{\bar\theta}\,\partial_\lambda
q_{\bar\theta}-q_{\bar\theta}\,\partial_\lambda{\bar\theta}}
{\frac{N}{c}\,q^2_{\bar\theta}+\frac{c}{N}\,p^2_{\bar\theta}}(\tau
,\lambda ),\nonumber \\
 &&\nonumber\\
 \partial_{\lambda}\, \bar \phi(\tau ,\lambda ) &=& \frac{p_{\bar\phi}\,
 \partial_\lambda q_{\bar\phi}-q_{\bar\phi}\,\partial_\lambda{\bar\phi}}
{\frac{N}{c}\,q^2_{\bar\phi}+\frac{c}{N}\,p^2_{\bar\phi}}(\tau
,\lambda ),
 \label{6.2}
 \eea

\noindent so that we have

\bea
 &&M = \int^{\pi}_o d\lambda\, \sqrt{-N^2 + A^2(\tau ,\lambda )+B^2(\tau ,\lambda )},
 \nonumber \\
 &&{}\nonumber \\
 &&\nonumber\\
 A(\tau ,\lambda ) &=& - {1\over 2}\, \frac
 {p_{\bar\theta}\partial_\lambda q_{\bar\theta}-q_{\bar\theta}\partial_\lambda p_{\bar\theta}}
 {\frac{N}{c}q^2_{\bar\theta}+\frac{c}{N}p^2_{\bar\theta}}(\tau ,\lambda
 )\, \int_o^\lambda d\lambda_1\,\Big(p_{\bar\theta}\partial_\lambda
 q_{\bar\theta} - q_{\bar\theta}\partial_\lambda p_{\bar\theta} +
 p_{\bar\phi}\partial_\lambda q_{\bar\phi}-q_{\bar\phi}\partial_\lambda p_{\bar\phi}
 \Big)(\tau, \lambda_1) - \nonumber\\
 &&\nonumber\\
 &-& \Big[\frac{N}{c}q_{\bar\theta}\partial_\lambda q_{\bar\theta}-
 \frac{c}{N}p_{\bar\theta}\partial_\lambda p_{\bar\theta}
 + \frac{N}{2c}\frac{q_{\bar\theta}}{p_{\bar\theta}}\Big(
 p_{\bar\phi}\partial_\lambda q_{\bar\phi} + q_{\bar\phi}\partial_\lambda p_{\bar\phi}
 \Big)\Big](\tau ,\lambda ), \nonumber\\
 &&\nonumber\\
 &&\nonumber\\
 B(\tau ,\lambda ) &=&
 \Big[\frac{p_{\bar\theta}}{2\, \sqrt{\frac{N^2}{c^2}q^2_{\bar\theta}+p^2_{\bar\theta}}}\,
 \frac {p_{\bar\phi}\partial_\lambda q_{\bar\phi}-q_{\bar\phi}\partial_\lambda p_{\bar\phi}}
 {\frac{N}{c}q^2_{\bar\phi}+\frac{c}{N}p^2_{\bar\phi}}\Big](\tau ,\lambda
 )\nonumber \\
 && \int_o^\lambda d\lambda_1\,\Big(
 p_{\bar\theta}\partial_\lambda q_{\bar\theta}-q_{\bar\theta}\partial_\lambda p_{\bar\theta}+
 p_{\bar\phi}\partial_\lambda q_{\bar\phi}-q_{\bar\phi}\partial_\lambda p_{\bar\phi}
 \Big)(\tau, \lambda_1) +\nonumber\\
 &&\nonumber\\
 &+& \Big[\frac{\sqrt{\frac{N^2}{c^2}q^2_{\bar\theta}+p^2_{\bar\theta}}}{p_{\bar\theta}}\,
 \Big(\frac{N}{c}q_{\bar\phi}\partial_\lambda q_{\bar\phi}+
 \frac{c}{N}p_{\bar\phi}\partial_\lambda p_{\bar\phi}\Big)
 +\frac{N}{c}\,\frac{q_{\bar\theta}}{p^2_{\bar\theta}}\Big(
 \frac{N}{2c}q^2_{\bar\phi}+\frac{c}{2N}p^2_{\bar\phi}
 \Big)\frac{p_{\bar\theta}\partial_\lambda q_{\bar\theta}-q_{\bar\theta}\partial_\lambda p_{\bar\theta}}
 {\sqrt{q^2_{\bar\theta}+\frac{c^2}{N^2}p^2_{\bar\theta}}}\nonumber\\
 &&\nonumber\\
 &-&\frac{1}{2}\,q_{\bar\theta}
 \frac{p_{\bar\phi}\partial_\lambda q_{\bar\phi}-q_{\bar\phi}\partial_\lambda p_{\bar\phi}}
 {\frac{N}{c}q^2_{\bar\phi}+\frac{c}{N}p^2_{\bar\phi}}
 \sqrt{\frac{N^2}{c^2}q^2_{\bar\theta}+p^2_{\bar\theta}}\Big](\tau ,\lambda
 ).
 \label{6.3}
 \eea
\medskip

We get the following spin vector [non-linear realization of O(3)]

\bea
 \vec S &=& \int^{\pi}_{o} d\lambda\, \vec S(\tau
 ,\lambda ),\nonumber \\
 &&{}\nonumber \\
 S^1(\tau ,\lambda ) &=&  \Big({{p_{\bar \phi}\, (\frac{N}{2c}q^2_{\bar \theta}
 + \frac{N}{2c}p^2_{\bar \theta})}\over {\sqrt{\frac{N^2}{c^2}q^2_{\bar \phi} + p^2_{\bar \phi}}}} +
 \left(\frac{N}{c}\right)^{3/2}
 {{q_{\bar \theta}\, q_{\bar \phi}\, \sqrt{\frac{N}{c}q^2_{\bar \phi} + \frac{c}{N}p^2_{\bar \phi}}}\over
 {2\,p_{\bar \theta}}}\Big)(\tau ,\lambda ),\nonumber \\
 S^2(\tau ,\lambda ) &=&  \Big({{q_{\bar \phi}\,
(\frac{N}{2c}q^2_{\bar \theta} + \frac{c}{2N}p^2_{\bar
\theta})}\over {\sqrt{q^2_{\bar \phi} + \frac{c^2}{N^2}p^2_{\bar
\phi}}}} - \sqrt{\frac{N}{c}} {{\,q_{\bar \theta}\, p_{\bar \phi}\,
\sqrt{\frac{N}{c}q^2_{\bar \phi} + \frac{c}{N}p^2_{\bar \phi}}}\over
 {2\,p_{\bar \theta}}}\Big)(\tau ,\lambda ),\nonumber \\
 S^3(\tau, \lambda) &=& -\left(\frac{N}{2c}q^2_{\bar \phi} + \frac{c}{2N}p^2_{\bar \phi}\right))(\tau ,\lambda ).
 \label{6.4}
 \eea
\bigskip

Eqs.(\ref{6.4}) are well defined either if $q_{\bar \theta}(\tau
,\lambda ) \not= 0$ for every $\lambda$ or if $p_{\bar \theta}(\tau
,\lambda )/ q_{\bar \theta}(\tau ,\lambda )$  tends to a finite
limit for $q_{\bar \theta}(\tau ,\lambda ) \rightarrow 0$.

\bigskip

Finally we can introduce a canonical basis $a_i(\tau ,\lambda )$,
$a^*_i(\tau ,\lambda )$, $i=1,2$, of quantities which are the
classical background for  two sets of creation and annihilation
operators

\bea
 q_{\bar \theta}(\tau, \lambda) &=& \sqrt{\frac{c}{2N}}\, (a_1 +
 a_1^*)(\tau, \lambda),\qquad p_{\bar \theta}(\tau, \lambda) = -
 i\, \sqrt{\frac{N}{2c}}\, (a_1 - a_1^*)(\tau, \lambda), \nonumber \\
  q_{\bar \phi}(\tau, \lambda) &=& \sqrt{\frac{c}{2N}}\, (a_2 +
  a_2^*)(\tau, \lambda),\qquad p_{\bar \phi}(\tau, \lambda) = -
  i\, \sqrt{\frac{N}{2c}}\, (a_2 - a_2^*)(\tau, \lambda), \nonumber \\
 a_1(\tau, \lambda) &=& \Big(\sqrt{\frac{N}{2c}}\,q_{\bar \theta} +
 i\,\sqrt{\frac{c}{2N}}\, p_{\bar \theta}\Big)(\tau, \lambda), \qquad
 a_2(\tau, \lambda) = \Big(\sqrt{\frac{N}{2c}}\,q_{\bar \phi} +
 i\,\sqrt{\frac{c}{2N}}\, p_{\bar \phi}\Big)(\tau, \lambda).
 \label{6.5}
 \eea
\medskip

In terms of these variables the invariant mass and the rest spin
have the following expression

 \begin{eqnarray*}
  M &=& \int^{\pi}_o d\lambda\, \sqrt{-N^2 + A^2(\tau ,\lambda )+B^2(\tau ,\lambda )},
 \nonumber \\
 &&{}\nonumber \\
 &&{}\nonumber\\
 A(\tau, \lambda) &=& \frac{a_1\partial_\lambda a_1^+-a_1^+\partial_\lambda a_1}
 {2\, \sqrt{2a_1^+a_1}}(\tau, \lambda)\,
 \int_o^\lambda d\lambda_1\,
 \Big(a_1\partial_\lambda a_1^+-a_1^+\partial_\lambda a_1+
 a_2\partial_\lambda a_2^+-a_2^+\partial_\lambda a_2\Big)(\tau,
 \lambda_1) + \nonumber\\
 &&\nonumber\\
 &+&\Big[a_1\partial_\lambda a_1^+ + a_1^+\, \partial_\lambda a_1
 -\frac{1}{2}\Big(a_2\partial_\lambda a_2^+ - a_2^+\, \partial_\lambda a_2\Big)
 \frac{a_1+a_1^+}{a_1-a_1^+}\Big](\tau, \lambda),
 \end{eqnarray*}

 \begin{eqnarray*}
 B(\tau, \lambda) &=& \frac{i}{4}\, \Big(\frac{a_1-a_1^+}{\sqrt{a_1^+a_1}}\,
 \frac{a_2\partial_\lambda a_2^+-a_2^+\partial_\lambda a_2}{\sqrt{2a_2^+
 a_2}}\Big)(\tau, \lambda)\nonumber \\
 &&\int_o^\lambda d\lambda_1\,
 \Big(a_1\partial_\lambda a_1^+-a_1^+\partial_\lambda a_1+
 a_2\partial_\lambda a_2^+-a_2^+\partial_\lambda a_2\Big)(\tau,
 \lambda_1) + \nonumber \\
 &&\nonumber\\
 &+& \Big[2i\,\frac{\sqrt{a_1^+a_1}}{a_1-a_1^+}\,\Big(
 a_1\partial_\lambda a_1^++a_1^+\partial_\lambda a_1
 \Big) -2i\, \frac{a_1^++a_1}{(a_1-a_1^+)2}\,\frac{a_2^+a_2}{\sqrt{a_1^+a_1}}\,
 \Big(a_1\partial_\lambda a_1^+-a_1^+\partial_\lambda a_1\Big)
 \nonumber\\
 &&\nonumber\\
 &+&i\,\frac{a_1+a_1^+}{2}\,\frac{\sqrt{a_1^+a_1}}{\sqrt{2a_2^+a_2}}
 \Big(a_2\partial_\lambda a_2^+-a_2^+\partial_\lambda a_2\Big)\Big](\tau,
 \lambda),\end{eqnarray*}

\bea
  \vec S &=& {1\over 2}\, \int^{\pi}_{- \pi} d\lambda\, \vec S(\tau
 ,\lambda ),\nonumber \\
 &&{}\nonumber \\
 S^1(\tau ,\lambda ) &=& - {i\over 2}\, {{(a_1^* - a_1)\,
 (a_2^* - a_2)\, a_1^*\, a_1 - (a_1^* + a_1)\, (a_2^* + a_2)\, a_2^*\, a_2}\over
 {(a_1^* - a_1)\, \sqrt{a_2^*\, a_2}}}(\tau ,\lambda ),\nonumber \\
 S^2(\tau ,\lambda ) &=& + {i\over 2}\, {{ (a_1^* - a_1)\,
 (a_2^* + a_2)\, a_1^*\, a_1 + (a_1^* + a_1)\, (a_2^* - a_2)\, a_2^*\, a_2}\over
 {(a_1^* - a_1)\, \sqrt{a_2^*\, a_2}}}(\tau ,\lambda ),\nonumber \\
 S^3(\tau ,\lambda ) &=& - \, \Big(a^*_2\, a_2\Big)(\tau, \lambda).
 \label{6.5}
 \eea

This representation is well defined if we have $a_1^*(\tau, \lambda)
- a_1(\tau, \lambda) \not= 0$ for every $\lambda$.

 \vfill\eject

\section{Conclusions}

In this paper we have developed the rest-frame instant form of the
open Nambu string (the closed one could be treated in the same way).
The string can be described as a frozen decoupled non-local
canonical non-covariant Newton-Wigner center of mass ($\vec z$,
$\vec h$) plus a canonical basis of Wigner-covariant relative
variables restricted by transversality first-class constraints
implying that only modes orthogonal to the string are physical. The
Newton-Wigner center of mass carries a universal external
realization of the Poincare' algebra, which depends on the string
relative variables only through the invariant mass $M$ and the
rest-spin ${\vec {\bar S}}$.\medskip

We have found two canonical bases of gauge invariant Dirac
observables by using Frenet-Serret geometrical methods and by
Abelianizing the transversality constraints.\medskip

The first canonical basis of Dirac observables is of the type of
action-angle variables and it can be reformulated in terms of two
sets of transverse oscillators, giving rise to a non-linear
realization of the O(3) algebra of the rest spin (like in a
$\sigma$-model).\medskip

The second one, where the Dirac observables are Lorentz scalars, is
only implicitly known. In it the Dirac observables are the
Frenet-Serret curvature and torsion of the string plus the conjugate
momenta.\medskip

Let us end with some comments on canonical quantization. In the
usual approach one first quantizes and then makes the reduction to
the physical states by imposing the quantum constraints expressed in
terms of Fock operators satisfying the Virasoro algebra. This is
done in the light-cone gauge and  only in the critical dimension
dimension $d = 26$ does one get a quantum realization of the
Poincare' algebra. There is a tachyon and a naive treatment of the
relativistic center of mass. The papers in Refs. \cite{8a,11a,24a}
(and \cite{12a} for the closed string) were attempts to find
Wigner-covariant classical oscillators for this approach.\medskip

Another stimulus to develop the rest-frame instant form came from
Ref.\cite{25a}, were the string was studied with the techniques of
the inverse scattering methods (Lax pairs). This allowed one to find
a recursive relation for building action-angle variables (with the
action ones globally defined) and then special oscillators giving
rise to rising Regge trajectories and without tachyons. The final
statement of this approach was that a certain family of motions of
the string without cusp singularities ("smooth strings") can be
quantized in dimension $d = 4$. However one has still to solve the
constraints at the quantum level.

\medskip

Instead in Refs.\cite{26a,27a,28a} there is the definition of an
infinite-dimensional algebra of gauge-invariant conserved charges
(including the Poincare' generators), the so-called Pohlmeyer-Rehren
constants of the motion. Again the center of mass is treated in a
naive way. These charges are quantized in dimension $d = 4$ with
methods of algebraic quantization by means of a deformation of the
algebra which is supposed to be without anomalies (a final proof is
still lacking). See the review in Ref.\cite{29a}, where a different
quantization in the framework of loop quantum gravity is tried.
However, not withstanding all the technical developments, it is
shown in Ref.\cite{30a} that this quantization is not equivalent to
the standard canonical quantization with a Fock space: in every
dimension there are anomalies if the invariant charges are quantized
in a Fock space.

\bigskip

The classical results of this paper may be relevant for the approach
"first reduce and then quantize", which generically is inequivalent
to the standard one.\medskip

The open problem with our approach concerns the quantization method.
Since action-angle variables cannot be quantized consistently due to
the angles \cite{31a}, one should quantize the non-linear
realization of the O(3) algebra plus the invariant mass. Due to the
complexity of the non-linear realization of O(3), this is a highly
non-trivial problem and any result would be welcome (for instance a
statement on the existence of critical dimensions). As shown in
Ref.\cite{32a} the canonical quantization of the frozen decoupled
canonical Newton-Wigner center of mass can be done in a way which
avoids the causality problems connected with the Hegerfeldt theorem
(instantaneous spreading of wave packets) and allows one to define
an intrinsic Wigner-covariant inner space of rest relative
variables. Moreover the non-locality of the non-covariant center of
mass makes it not observable (like the wave function of the
universe!) and this makes its universal non-covariance irrelevant.
If $M$ and $\vec S$ can become self-adjoint operators with the spin
operator satisfying the O(3) algebra and commuting with the
invariant mass operator then there is no problem in quantizing the
external Poincare' algebra.

\medskip

The same scheme should be used if we could express the invariant mass and the rest
spin in terms of the FS curvature and torsion and of the conjugate
momenta. However, since we have $\rho_1(\tau ,\lambda ) \geq 0$ and
$\rho_2(\tau ,\lambda ) \geq 0$, these variables are like radial
variables and in trying to quantize them one should use the
non-canonical basis $r$ and $r\, p_r$, the only one which would give
meaningful commutators due to $r \geq 0$.

\vfill\eject

\appendix

\section{Distributions for the Nambu String}

We collect here the definitions of the  most important distributions
used in this paper, which were given in the Appendix of
Ref.\cite{11a}. Many of them where already present in
Ref.\cite{19a}.\medskip

The even and odd $\delta$-functions with  $2\pi$-periodicity are
respectively

\bea
 \Delta_+(\sigma ,\bar \sigma)&=& \frac{1}{2\, \pi}\,
\sum_{n = - \infty}^{\infty}\, \Big[ e^{i\, n\, (\sigma - \bar
\sigma)} + e^{i\, n\, (\sigma + \bar \sigma)} \Big] = \sum_{n = -
\infty}^{\infty}\, \Big[ \delta\, (\sigma - \bar \sigma + 2\, n\,
\pi) + \delta\, (\sigma + \bar \sigma + 2\, n\, \pi) \Big]
=\nonumber \\
 &=& \partial_{\sigma}\, \sum_{n = - \infty}^{\infty}\, \Big[
 \theta (\sigma - \bar \sigma + 2\, n\, \pi) + \theta
(\sigma + \bar \sigma + 2\, n\, \pi) \Big] = \frac{1}{\pi} +
\partial_{\sigma}\, \Sigma_+(\sigma , \bar \sigma ),
 \label{a1}
 \eea

\noindent and

\bea
 \Delta_-(\sigma , \bar \sigma) &=& \frac{1}{2\, \pi}\,
 \sum_{n = - \infty}^{\infty}\, \Big[e^{i\, n\, (\sigma - \bar \sigma)}
- e^{i\, n\, (\sigma + \bar \sigma)} \Big] = \sum_{n = - \infty
}^{\infty}\, \Big[ \delta\, (\sigma - \bar \sigma + 2\, n\, \pi) -
\delta\, (\sigma + \bar \sigma + 2\, n\, \pi) \Big]=\nonumber \\
 &=& \partial_{\sigma}\, \sum_{n = - \infty}^{\infty}\, \Big[
 \theta(\sigma - \bar \sigma + 2\, n\, \pi) - \theta
(\sigma + \bar \sigma + 2\, n\, \pi) \Big] =
\partial_{\sigma}\, \Sigma_-(\sigma , \bar \sigma ) = - \partial_{\bar \sigma}\,
 \Sigma_+(\sigma , \bar \sigma ),\nonumber \\
 &&{}
 \label{a2}
 \eea

\noindent where $\theta (\sigma )$ is the step function ($= 1$ for
$\sigma > 0$; $= 0$ for $\sigma < 0$; $\theta ( \sigma ) + \theta(-
\sigma ) = 1$) and $\delta\, (\sigma )$ $=
\partial_{\sigma}\, \theta (\sigma )$ is the standard $\delta$-function. The sign-function
is $\varepsilon (\sigma )$ $= \theta (\sigma ) - \theta (- \sigma)$
($= 1$ for $\sigma > 0$; $= - 1$ for $\sigma < 0$; $\varepsilon
(\sigma ) + \varepsilon (- \sigma) = 0$).\medskip

The delta functions  $\Delta_{\pm}$ have the following properties

\bea
 &&\Delta_+(\lambda, \lambda') = \Delta_+(- \lambda, \lambda') =
\Delta_+(\lambda', \lambda) = \Delta_+(\lambda + 2\, n\, \pi,
\lambda'),\nonumber \\
 &&\Delta_-(\lambda, \lambda') = - \Delta_-(- \lambda, \lambda') =
 \Delta_-(\lambda', \lambda) = \Delta_-(\lambda + 2\, n\, \pi, \lambda'),
 \nonumber \\
  &&\Delta^{'}_{\pm}(\lambda ,\lambda') = \displaystyle{\partial\
\over{\partial\lambda}}\, \Delta_{\pm}(\lambda, \lambda') = -
\displaystyle{\partial\  \over{\partial\lambda'}}\,
\Delta_{\mp}(\lambda,\lambda'),\nonumber \\
 &&\int_{- \pi}^{\pi}\, d\lambda'\, f(\lambda')\, \Delta_{\pm}(\lambda', \lambda) =
f(\lambda) \pm f(-\lambda).
 \label{a3}
 \eea
 \medskip

We give some formulas for the distributions $\Delta_{\pm}(\sigma,
\sigma\pri)$, used in the evaluation of some Poisson brackets. If
$f(x)$ and $g(x)$ are periodic functions, with period $2\, \pi$, and
with definite parity given by $P_{f} = \pm 1,\qquad {\rm if} \qquad
f(x) = \pm f(-x),$ where, of course, $P_{f'} = - P_{f}$, we have the
following identities

\bea
 &&f(x)\, g(y)\, \Delta_{\pm}^{\prime}(x,y) = f(y)\, g(y)\, \Delta^{\prime} _{\pm
 P_f}(x,y) - f^{\prime}(x)\, g(x)\, \Delta_{\pm P_g}(x,y),\nonumber \\
 &&f(x)\, g(y)\, \Delta^{\prime}_+(x,y) \pm f(y)\, g(x)\, \Delta^{\prime}
 _-(x,y) = \pm f(x)\, g(x)\, \Delta^{\prime}_{-P_f}(x,y)+\nonumber \\
 &&+ f(y)\, g(y)\, \Delta^{\prime}_{+P_f}(x,y) - f^{\prime}(x)\, g(x)\, \Big[ \Delta
_{+P_g}(x,y) \mp \Delta_{-P_f}(x,y)\Big],\nonumber \\
 &&\Big[f(x)\, g(y) - f(y)\, g(x)\Big]\, \Delta^{\prime}_{\pm}(x,y)
= \Big[f(y)\, g(y) - f(x)\, g(x)\Big]\, \Delta^{\prime}_{\pm
P_f}(x,y)-\nonumber \\
 &&- f^{\prime}(x)\, g(x)\, \Big[\Delta_{\pm
P_g}(x,y) + \Delta_{\pm P_f}(x,y) \Big],\nonumber \\
 &&\Big[f(x)\, g(y) + f(y)\, g(x)\Big]\, \Delta^{\prime}_{\pm}(x,y)
= \Big[f(y)\, g(y) + f(x)\, g(x)\Big]\, \Delta^{\prime}_{\pm
P_f}(x,y)-\nonumber \\
 &&- f^{\prime}(x)\, g(x)\, \Big[\Delta_{\pm
P_g}(x,y) - \Delta_{\pm P_f}(x,y) \Big].
 \label{a11}
 \eea

\vfill\eject

\end{document}